\documentstyle[11pt,twoside,epsfig]{article}
\newcommand{\pom}{{I\!\!P}}

\newlength{\dinwidth}
\newlength{\dinmargin}
\begin{document}
\begin{titlepage}
\begin{flushleft}
%
%
{\tt DESY 96-122    \hfill    ISSN 0418-9833 \\
  June 1996}   
\end{flushleft}
\vspace*{2.cm}
%
%
\vspace{1cm}
\begin{center}
\begin{Large}
{\bf
Strangeness Production in Deep-Inelastic Positron-Proton \\
Scattering at HERA.}  \\
\vspace*{2.cm}
H1 Collaboration \\
\vspace*{1.cm}
\end{Large}
\vspace*{1.cm}
{\bf Abstract}
\begin{quotation}
\noindent
Measurements are presented of $K^0$ 
meson and $\Lambda$ baryon production in 
deep-inelastic positron-proton scattering (DIS)
in the kinematic range $10 < Q^2 < 70\,$GeV$^2$ and
$10^{-4} < x < 10^{-2}$.
The measurements, obtained using the 
H1 detector at the HERA
collider, are discussed
in the light of possible mechanisms for increased 
strangeness production at low Bjorken-$x$.   
Comparisons of the $x_F$ spectra,
where $x_F$ is the fractional longitudinal 
momentum in the hadronic centre-of-mass frame,
with results from electron-positron 
annihilation are made. 
The $x_F$ spectra
and the $K^0$ ``seagull'' plot are compared with previous DIS results.
The mean $K^0$ and $\Lambda$ multiplicities
are studied as a function of the centre-of-mass energy $W$ and  
are observed to be consistent with 
a logarithmic increase with 
$W$ when compared with previous
measurements. A comparison of the levels of
strangeness production in diffractive and 
non-diffractive DIS is made.
An upper limit of $0.9\,$nb, at the $95\%$ confidence level,
is placed on the cross-section for QCD instanton
induced events.  
\end{quotation}
\vfill
%
%
\cleardoublepage
\end{center} \end{titlepage}
\begin{Large} \begin{center} H1 Collaboration \end{center} \end{Large}
\begin{flushleft}

 S.~Aid$^{13}$,                 
 M.~Anderson$^{23}$,            
 V.~Andreev$^{26}$,             
 B.~Andrieu$^{29}$,             
 R.-D.~Appuhn$^{11}$,           
 A.~Babaev$^{25}$,              
 J.~B\"ahr$^{36}$,              
 J.~B\'an$^{18}$,               
 Y.~Ban$^{28}$,                 
 P.~Baranov$^{26}$
% ...

\bigskip{\it
% ...
$ ^{2}$ III. Physikalisches Institut der RWTH, Aachen, Germany$^{ a}$ \\
% ...
$ ^{11}$ DESY, Hamburg, Germany$^{ a}$ \\
% ...
$ ^{13}$ II. Institut f\"ur Experimentalphysik, Universit\"at Hamburg,
          Hamburg, Germany$^{ a}$ \\
% ...
$ ^{18}$ Institute of Experimental Physics, Slovak Academy of
          Sciences, Ko\v{s}ice, Slovak Republic$^{ f}$ \\
% ...
$ ^{23}$ Physics Department, University of Manchester,
          Manchester, UK$^{ b}$ \\
% ...
$ ^{25}$ Institute for Theoretical and Experimental Physics,
          Moscow, Russia \\
 $ ^{26}$ Lebedev Physical Institute, Moscow, Russia$^{ f}$ \\
% ...
$ ^{28}$ LAL, Universit\'{e} de Paris-Sud, IN2P3-CNRS,
          Orsay, France \\
 $ ^{29}$ LPNHE, Ecole Polytechnique, IN2P3-CNRS, Palaiseau, France \\
% ...
$ ^{36}$ DESY, Institut f\"ur Hochenergiephysik, Zeuthen, Germany$^{ a}$ \\
% ...
$ ^a$ Supported by the Bundesministerium f\"ur Bildung, Wissenschaft,
      Forschung und Technologie, FRG,
      under contract numbers 6AC17P, 6AC47P, 6DO57I, 6HH17P, 6HH27I,
      6HD17I, 6HD27I, 6KI17P, 6MP17I, and 6WT87P \\
 $ ^b$ Supported by the UK Particle Physics and Astronomy Research
      Council, and formerly by the UK Science and Engineering Research
      Council \\
% ...
$ ^f$ Supported by the Deutsche Forschungsgemeinschaft \\
% ...
}
\end{flushleft}
%
\newpage

\section{Introduction}
This paper describes studies of the flavour structure of the 
hadronic final state in deep-inelastic, neutral current, 
positron-proton scattering (DIS) performed using the H1 
detector at the HERA collider at DESY. 
The scattering process is described in terms of the 
variables $Q^2=-q^2$, where $q$ is the four-momentum of the exchanged
boson, $x=Q^2/2p\cdot q$ and $y=p\cdot q/p\cdot k$
where $p$ and $k$ are, respectively, the four-momenta of the
initial state proton and electron.
Results from HERA have shown that the 
proton structure function $F_2(x,Q^2)$ 
measured in DIS rises rapidly 
as the Bjorken variable $x$ decreases below about 
$x=10^{-2}$~\cite{F295,ZF295}. 
Here
it is examined if this rise is accompanied by a change in the flavour
composition of the hadronic final state,
in particular the strangeness production rate, associated
with either the hard interaction or the subsequent hadronisation process. 

Typically in
electron-positron annihilation and 
DIS experiments at moderate or large $x$
the production ratios of $u$, $d$, and $s$ 
quarks in the soft hadronisation process have
been found to be $1:1:0.3$~\cite{Oldees2d,Olddiss2d}, 
that is $\lambda_s = 0.3$, where $\lambda_s$
is the strangeness suppression factor,
the ratio of $s$ quark to $u$ or to $d$
quark production necessary to describe the data using the 
Lund string hadronisation model~\cite{Lund}.
The partial suppression of $s$ quark
production in the 
hadronisation is a consequence of the relative masses
of the $u$, $d$ and $s$ quarks; the mass of the $c$ quark is 
such that $c$ quarks are not produced in the hadronisation process. 
Most recent DIS results~\cite{E665,ZK0,WA21b} favour 
a suppression of strangeness production stronger than that above,
$\lambda_s \approx 0.2$, as does one older measurement~\cite{WA21a},
although there is a recent result which prefers the higher value
of $\lambda_s$~\cite{E632}. 
Recent electron-positron annihilation
measurements~\cite{DELPHI} also tend to 
favour $\lambda_s \approx 0.2$, but again the 
experimental situation is not clear~\cite{ALEPH}.

In hadron-hadron collisions the mean value of the 
strangeness suppression factor
is about $0.3$ but there is evidence for a dependence
on the centre-of-mass energy, the proportion of strangeness
produced rising with energy. There is also some evidence 
for a dependence on the region of phase space 
investigated~\cite{Wroblewski}.
 
HERA offers the possibility of studying strangeness 
production in situations similar to those
prevailing in all the above processes. 
Hadron production in DIS in the 
phase space region associated with the struck quark 
is expected to resemble
that in electron-positron annihilation.
This is confirmed by recent measurements~\cite{H1breit,Zbreit}.
The rapidity region between the struck quark and the proton remnant
in DIS is likely to be similar to the central plateau in 
hadron-hadron collisions. A recent comparison of hadron production
in photoproduction and deep-inelastic electron-proton 
scattering supports this possibility~\cite{DISphoto}. 
The proton remnant is likely to be similar to 
the hadron remnants in hadron-hadron scattering.
The phase space region in which strange particles may be identified
using the current H1 apparatus corresponds to the struck quark and 
central plateau regions. 
The broad agreement of the $\lambda_s $ results obtained in 
electron-positron annihilation with the measurements in the
struck quark and central plateau regions in DIS and the 
central plateau in hadron-hadron collisions  
suggests that the
colour forces responsible for hadron production 
in these regions are similar. 
It is of interest to examine whether this similarity persists
at the lowest values of $x$ accessible at HERA. 

A possible source of strangeness production levels 
in DIS at HERA above those expected from previous 
measurements is QCD instantons~\cite{Instants}.
Instantons are expected features of both QCD and the 
Glashow-Weinberg-Salam (GWS) model. 
While the cross-section for instanton induced processes 
is subject to large uncertainties, the properties of the final 
states of instanton induced events may be calculated. These include 
exciting signatures such as baryon plus lepton number 
non-conservation in
the case of the GWS model and, 
analogously, non-conservation of chirality
in the case of QCD.
The latter 
signature is difficult to observe experimentally, but a further 
property of QCD instanton 
induced events, the democratic production of all
kinematically allowed flavours, 
would lead to a measureable increase in the quantity of strangeness
in the final state. The kinematics of instanton induced events 
is such that this excess strangeness would be produced
largely in the regions of the H1 detector close to the
proton remnant.  

A further possible mechanism for the production of strangeness
at levels above expectations based on previous measurements 
may come into play at low $x$. Although the behaviour of $F_2(x,Q^2)$ in
this region can be accurately described within the framework of 
perturbative QCD using the Dokshitzer, Gribov, Lipatov, Altarelli and Parisi 
(DGLAP) evolution equations~\cite{DGLAP}, it has been argued that
the sharp rise in $F_2(x,Q^2)$ with decreasing $x$
is an indication that new QCD dynamics, described
by the Balitsky, Fadin, Kuraev and Lipatov (BFKL) 
evolution equation~\cite{BFKL}, is starting to play a role
at the lowest values of $x$ accessible to HERA~\cite{BFKLfit}.
If the BFKL description is appropriate 
it is expected that the hadronic final state 
will also exhibit features associated with BFKL evolution,
in particular the production of an increased number 
of gluons with large momentum transverse to the proton in the
region around the proton remnant direction~\cite{BFKLEt}. There 
are some indications that this is indeed observed~\cite{ForwJet}.
This would then imply an increased level of particle, including
strange particle, production in this region.
Indeed, the proportion of strange particles observed might also 
increase as there is some evidence 
that events containing energetic gluons in
electron-positron annihilation contain a higher
proportion of $K^0$ mesons than is observed in 
two-jet events~\cite{Deljetk}.

Studies at HERA have revealed that about $10\%$ of DIS 
events are diffractive in nature~\cite{F2D(3),ZF2D(3)}. 
These events exhibit a large 
region in rapidity, close to the proton remnant, in which 
there is no significant hadronic energy. 
It has been shown that a large proportion of them
may be described in terms of 
deep-inelastic positron-pomeron scattering. Furthermore, 
measurements are in agreement with Monte Carlo simulations in 
which the pomeron is assumed to be an object with partonic 
sub-structure and suggest that the pomeron is 
composed predominantly of gluons~\cite{Paris95,Zpomglue}. 

Describing the kinematics of 
diffractive deep-inelastic scattering (DDIS) using 
the positron-pomeron scattering picture
requires the introduction 
of two further variables:
$t$, the square of the four-momentum of 
the pomeron and $x_\pom$, the proportion of the proton's
four-momentum carried by the pomeron. The latter
may be determined using the
relationship
$$x_\pom \approx \frac{M_X^2 + Q^2}{W^2+Q^2},$$
where $M_X^2$ is the squared invariant mass of the photon-pomeron
system and $W^2$ is the squared invariant mass of the
complete hadronic final state. 
The approximation is good as long as
both the squared proton mass and
$|t|$ are much smaller than either $Q^2$ or $W^2$, 
which is the case in the kinematic region considered here.
In this picture of DDIS,
as the pomeron has vacuum quantum numbers, 
the production of flavour is expected to 
be democratic up to effects arising from the quark masses.

An alternative model for DDIS is based on the 
idea that the underlying process is electron-gluon 
scattering via boson-gluon fusion (BGF)
followed by the randomisation 
of the colours of the produced quark and
anti-quark~\cite{Buch}. The probability of this 
producing a colour singlet configuration is $\frac{1}{9}$.
In this case there is no colour connection between the 
proton remnant and the quark and anti-quark,
no hadrons are produced in this region, 
and the above diffractive signature is observed. 
In this model quark flavour production is
again democratic up to kinematic effects.
It is of interest to examine whether the quantity of
strangeness produced in DDIS events
is compatible with these models of diffraction.    

Also of interest is the
possibility that particles containing
strange quarks produced in the hard interaction may be identified.
Possible interaction mechanisms are:   
BGF resulting in the production of an 
$s\overline{s}$ pair, 
$c\overline{c}$ production
via BGF followed by decay to $s$ quarks,
and the direct interaction 
of the virtual boson with a strange sea quark. 
This may ultimately allow  
measurement of the gluon and strange sea contributions
to proton structure.

The above considerations
provide the motivation for the work reported here. 
Strangeness production is examined by measuring the numbers 
and proportions of
$K^0_S$ mesons and $\Lambda$ and $\overline{\Lambda}$
baryons present in the final state 
of DIS interactions. In the following, 
$K^0$ is taken to 
refer to both $K^0$ and $\overline{K^0}$ and $\Lambda$ 
to both $\Lambda$ and $\overline{\Lambda}$ unless
explicitly stated to the contrary. 

\section{The Detector}
A detailed description of the H1 detector can be 
found in~\cite{H1NIM}.
The following briefly describes the components 
most important to this analysis.
The coordinate system used is defined to have 
its origin at the nominal interaction point. The $z$ axis is 
chosen to be along the 
proton beam direction, the $y$ axis vertical and the $x$ axis 
completes the right-handed coordinate system. Where polar 
coordinates are used, $\theta$ is the angle with respect to the $z$
axis and $\phi$ the azimuthal angle with respect to the $x$ axis.
Forward and backward refer to the positive and negative $z$
directions respectively, hence
positive pseudo-rapidities $\eta=-\ln\tan(\theta/2)$ refer to objects
in the forward hemisphere. 

Surrounding the positron-proton interaction region are
detectors designed to measure the tracks followed by
charged particles. 
These are
the central track detector (CTD) 
which covers the polar angular range 
$15^\circ < \theta < 165^\circ$ ($2.0 > \eta > -2.0$)
and the forward track detector (FTD) covering the
range $7^\circ < \theta < 25^\circ$ ($2.8 > \eta > 1.5$).
The CTD consists of concentric
inner and outer jet drift 
chambers.
Located at the inner radii of these chambers are multi-wire  
proportional chambers (MWPCs), used for 
triggering purposes, and further 
drift chambers designed to measure accurately the $z$ position 
of charged particle tracks.
A uniform
magnetic field of $1.15\,$T parallel to the beam axis 
is maintained 
throughout the region containing the CTD and FTD
by a superconducting solenoid (see below).
This system enables the 
transverse momentum of a charged particle to be 
measured with a resolution of
$\sigma_{p_T}/p_T \approx 0.009 p_T\oplus0.015$
in the CTD ($p_T$ in GeV/c). 

The liquid argon (LAr) calorimeter~\cite{LArC} 
extends over the polar angular range
$4^\circ < \theta <  154^\circ$ ($3.4 > \eta > -1.5$)
and covers the entire azimuth.
It consists of an electromagnetic section with
lead absorbers, corresponding to a depth of between $20$ 
and $30$ radiation
lengths, and a hadronic section with steel absorbers.
The total depth of the calorimeter varies between $4.5$ and $8$
hadronic interaction lengths.

The  backward electromagnetic calorimeter~\cite{BEMC} (BEMC), 
which is $21.7$ radiation lengths deep, covers the  
region $151^\circ < \theta < 176^\circ$ ($-1.4 > \eta > -3.4$).
A resolution of $0.10/\sqrt{E}$ with $E$ in GeV 
has been achieved for positrons.
The BEMC energy scale is known to an accuracy of $1.7\%$.
Immediately in front of the BEMC is the backward proportional 
chamber (BPC) which determines the point at which the 
scattered positron enters the BEMC to an accuracy of $1.5\,$mm over the 
polar angle range $155^\circ < \theta < 174^\circ$ 
($-1.5 > \eta > -2.9$).
In conjunction with a measurement of the $z$ coordinate 
of the primary interaction this leads to 
an accuracy of $1\,$mrad for the measurement of the polar angle
of the scattered positron.
 
Behind the BEMC is a time-of-flight system (ToF) with 
a time resolution of 
about $1\,$ns. This enables the rejection of 
background events caused by the 
interaction of protons in the material before the H1 detector, as 
the particles from these events arrive at the ToF system earlier
than those from interactions at the nominal vertex.
 
The LAr calorimeter is surrounded by a superconducting 
solenoid which provides the 
magnetic field in the region containing the track detectors.

Behind the LAr calorimeter 
and the BEMC is an iron/streamer tube calorimeter which 
serves as the return yoke for the magnetic field and 
detects hadronic energy leaking out of the inner calorimeters. 
The iron/streamer tube system also serves as a muon detector.

Surrounding the beam-pipe in the forward direction 
is a copper/silicon calorimeter, the Plug.
This device covers the
range $0.7^\circ < \theta < 3.5^\circ$ ($5.1 > \eta > 3.5$).

Outside the iron in the forward direction is a muon spectrometer
which consists of a toroid magnet surrounding the beam-pipe 
sandwiched between two layers of drift chambers. 
In addition to its primary function of 
detecting forward-going muons this device 
detects tracks from the  
secondary interactions of particles in the  
range $0.2^\circ < \theta <  0.8^\circ $ ($6.6 > \eta > 5.0$).
These occur 
in the collimators designed to shield the detector from synchrotron
radiation, the forward beam-pipe and the adjoining material.

The luminosity measurement system consists of electromagnetic 
calorimeters placed at $z=-33\,$m and $z=-103\,$m~\cite{Lumi}.

\section{Monte Carlo Models}
Calculations of the expected rate of strangeness production 
in DIS are available in the form of Monte Carlo 
event generators. These
are used here, together with a Monte Carlo simulation
of the H1 detector,  
both to correct the measurements for the finite
acceptance and resolution of the detector and to 
make comparisons with theoretical predictions. 
The standard DIS Monte Carlo 
generators used
may be split into two classes. 
The first class consists of those expected to describe 
the hadronic final state in non-diffractive DIS 
(NDDIS), but which contain no explicit
modelling of diffractive DIS. The second class 
is designed explicitly to model DDIS events.
Here only one such generator is considered.

The NDDIS generators are 
LEPTO 6.1~\cite{LEPTO}, 
ARIADNE 4.03~\cite{ARIADNE} and HERWIG 5.8~\cite{HERWIG}. 
These generators simulate
positron-proton collisions using 
the leading order electroweak matrix element for the interaction
of a virtual boson with a parton, including 
exact leading order QCD corrections. The parton distributions used are 
the MRS(H) set of the Durham group~\cite{Durham} evolved to the
appropriate $Q^2$ using the DGLAP formalism. 
These parton distributions were determined
using most relevant available data, including the 
early HERA measurements of the 
proton structure function $F_2(x,Q^2)$~\cite{F292,ZF292}.
The subsequent calculation 
differs in the three generators.
LEPTO simulates higher order perturbative QCD effects using 
independent parton showers in the leading logarithm approximation 
to generate radiation from
the partons entering and leaving the hard sub-process.
The observed hadrons are then produced 
using the Lund string hadronisation model~\cite{Lund}, as implemented
in JETSET 7.4~\cite{JETSET}.
Three sets of fragmentation
parameters are used 
in JETSET in this study: the default 
values which give a good 
description of PETRA and PEP 
electron-positron annihilation data,
those providing the best description
of DELPHI electron-positron annihilation
data~\cite{DELPHI} and those which best describe 
E665 muon-nucleon scattering data~\cite{E665}.
These parameters are listed in 
table~\ref{tab001}.
\begin{table}
\begin{center}
\begin{tabular}[hbpt]{||l|l|c|c|c||}
\hline 
Parameter & Description  &  Default value & DELPHI value & E665 value \\
\hline       
Parj(2) & $\lambda_s $ & $0.3$ & $0.23$ & $0.2$ \\ 
\hline       
Parj(11) & Prob. meson has spin 1 ($u$, $d$) 
& $0.5$ & $0.365$ & $0.5$ \\ 
\hline       
Parj(12) & Prob. meson has spin 1 ($s$) 
& $0.6$ & $0.410$ & $0.6$  \\ 
\hline        
\end{tabular}
\caption{Values of some JETSET hadronisation parameters
         associated with the production of $u$, $d$ and $s$ quarks
         required to give a good description of 
         PETRA and PEP data, DELPHI and E665 data.}
  \label{tab001}
\end{center}
\end{table}
The term CDM
is used to refer to
a calculation in which higher order perturbative QCD effects  
are simulated using ARIADNE, an implementation of the 
Colour Dipole Model~\cite{CDM}. In this model gluon radiation
takes place initially from the dipole formed by the colour charges 
separated by the hard scattering process and subsequently 
from the dipoles formed by the radiated gluons. 
Again hadronisation
is done using JETSET with the above parameter sets.
HERWIG simulates parton showers including both colour coherence and 
soft gluon interference effects. Hadronisation is simulated 
by the non-perturbative splitting of 
the gluons generated in the parton 
showers to create colour neutral clusters,
followed by the fragmentation of those clusters.

In addition to the above, DJANGO 6.0~\cite{DJANGO} 
is used to study the
effects of initial state photon radiation. This Monte Carlo generator 
simulates the electroweak interaction using 
HERACLES~\cite{HERACLES} and the hadronic final state using LEPTO.

The DDIS Monte Carlo used  
is RAPGAP 2.1~\cite{RAPGAP}.
In this generator a pomeron flux, dependent only on $t$ and $x_\pom$, 
is associated with the proton. 
The pomeron is assumed to have partonic 
structure which evolves with $Q^2$ according to the
DGLAP prescription.  
The virtual boson interacts with a parton within the
pomeron, the matrix element used being as for the NDDIS
Monte Carlo generators. Further parton emission is 
simulated using either parton showers
(RAPGAP-PS) or the 
Colour Dipole Model (RAPGAP-CDM) and hadronisation is 
performed using JETSET. In conjunction with RAPGAP
only the default fragmentation parameters
are used. 
The pomeron parton distributions chosen give 
a good description of the latest
H1 measurements of $F_2^{D(3)}$~\cite{Moriond}.

\begin{figure}[htbp]  \centering
\epsfig 
 {file=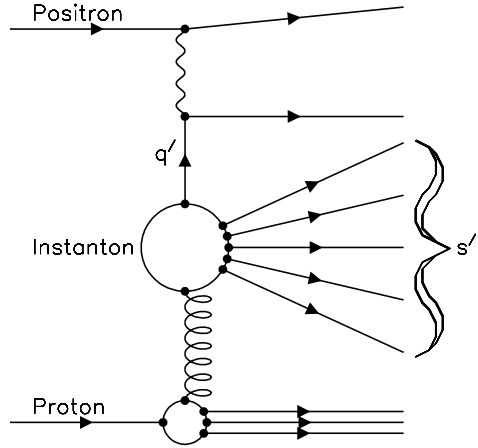,height=80mm,width=80mm,angle=0}
\caption{Diagram illustrating the kinematics
of instanton induced deep-inelastic positron-proton scattering.}
  \label{fig001}
\end{figure}

Calculations of the expectations for the 
hadronic final state of instanton induced processes are
made using QCDINS~\cite{QCDINS}, which is a version of HERWIG 
modified to generate QCD instanton induced events. The number of 
flavours generated in the instanton decay is determined by the
available instanton sub-process centre-of-mass energy, $\sqrt{s'}$. 
The decay is assumed to 
be isotropic in that frame.
The minimum value of the 
so-called Holy Grail function~\cite{Instants}, 
which enters the expression for
the cross-section for instanton induced events, is 
fixed to be the value it reaches at
$x'=0.2$, where $x'=Q'^2/(s'+Q'^2)$ and 
$Q'^2=-q'^2$ is the negative invariant mass of the
virtual quark in the instanton sub-process, see 
figure~\ref{fig001}.

The number of Monte-Carlo events used in comparisons with the  
measurements presented in the following always exceeds the 
number of events in the data by a factor of three or more. 
 
\section{Data Taking and Event Selection}
The data used here were collected in the 1994 running period in 
which $820\,$GeV protons and $27.5\,$GeV positrons
were brought into head-on collision in HERA. 

The trigger used in the collection of the data is the requirement
that there be a localised energy deposit in the BEMC of at least
$4\,$GeV with no associated veto signal from the ToF system.
This has been shown to be more than $99\%$ efficient for the
collection of DIS events in which the scattered positron 
enters the BEMC and has energy greater than
$10\,$GeV.
 
The events are then required to contain a 
localised BEMC energy deposit 
of at least $12\,$GeV 
spatially associated with a signal in the
BPC and with a shape compatible with being due to an electromagnetic 
shower. It is also demanded that at least one charged particle be
detected in the CTD, allowing the determination of the
$z$ coordinate of the primary interaction vertex. This is required
to be within $30\,$cm of the $z$ coordinate of the expected
vertex position. 

The polar angle of the positron candidate is
required to lie in the range $156^\circ < \theta < 173^\circ$.
The range in the kinematic variables studied is 
restricted to $10 < Q^2 < 70\,$GeV$^2$, 
$10^{-4} < x < 10^{-2}$
and $0.05 < y < 0.6$, all variables being 
determined using the measured 
polar angle and energy of the scattered positron. The 
lower restriction in
the $y$ range ensures that these determinations are of sufficient 
accuracy. 
The above criteria are such that the boundaries of the kinematic
region studied are given by the restrictions imposed on the 
Lorentz invariant variables $Q^2$, $x$ and $y$.
The number of events which satisfy the above is $53\,360$. 
Less than $1\%$ of these are background events. The largest 
source of background is
photoproduction interactions in which 
one of the final state hadrons fakes the required BEMC electron
signal. 

The luminosity to which the data correspond 
is $1.32\,$pb$^{-1}$, 
determined to a precision 
of $1.5\%$
by studying
the reaction $e p \rightarrow e p \gamma$ 
using the luminosity measurement system. 

Similar criteria 
to those used to isolate DDIS events in the study of 
the structure of deep-inelastic 
diffraction~\cite{F2D(3)} are used
here to identify diffractive events. These are:
the energy in the Plug must be less than $3\,$GeV;
there must be fewer than 3 hit pairs in the
forward muon drift chambers and
the most forward significant energy deposit detected in the LAr
calorimeter 
must be at a pseudo-rapidity of $3.0$ or less.
Events in which the forward-going proton remnant is separated 
from the centrally produced hadronic system by a rapidity gap
between $\eta \sim 6.6$ and $\eta = 3.2$
and has a mass, $m_R$, less than 
$1.6\,$GeV/c$^2$ satisfy these criteria.
The DDIS event sample is restricted
to the kinematic region defined by $x_\pom < 0.05$ and
contains $3\,491$ events. The variable $x_\pom$ 
is calculated from the values of $Q^2$ and $W^2$ determined using the 
parameters of the scattered electron and the value of $M_X$ obtained from 
the measurement of the hadronic final state.

As comparisons are made with 
Monte Carlo simulations which do not 
explicitly model the diffractive component of DIS, additional 
criteria are applied to select a sample of  
non-diffractive DIS events. 
These are that the events do not satisfy the diffractive selection 
above and in addition that the 
energy in the LAr calorimeter in the angular
region $4.4^\circ < \theta < 15^\circ$ ($3.26 > \eta > 2.03$) 
exceeds $0.5\,$GeV. 
The number of events in the NDDIS sample is $46\,684$.
The non-diffractive Monte Carlo simulations
are expected to model the hadronic
final state of these events. To ensure that any events
produced by these simulations which fall into the 
DDIS category are removed, before comparisons are made with
NDDIS data the Monte Carlo events are required to have a hadronic 
energy of at least $0.5\,$GeV in the above angular region.
\section{$K^0_S$ and $\Lambda$ Identification}
Neutral kaons are identified through the decay
$$K^0_S \rightarrow \pi^+ \pi^-,$$
and $\Lambda$ baryons through the decay
$$\Lambda \rightarrow p \pi^-$$ 
and the charge conjugate reaction. The identification 
utilises the
separation of the decay and primary vertices arising from 
the relatively long $K^0_S$ and $\Lambda$ lifetimes.

The first stage in the identification of
candidates for the above decays
is a search for
pairs of oppositely charged tracks in the CTD 
which originate from 
a vertex radially separated from the 
primary event vertex by at least $2\,$cm.  
Only well-measured
tracks with $p_T > 0.15\,$GeV/c are considered. 
An attempt is 
made to fit these track pairs using the hypothesis that they
originate from 
a common secondary
vertex resulting from the decay of a neutral particle, 
generically termed a $V^0$,
which in turn originates from the primary event vertex.
The $V^0$ candidates which satisfy this hypothesis
are then required to have 
transverse momenta in the 
range $0.25 < p_T^2 < 4.5\,($GeV/c$)^2$ and  
pseudo-rapidities in the range $-1.3 < \eta < 1.3$ to ensure that 
they are well within the acceptance of the CTD and that the
$V^0$ mass resolution is good. 

In order to define a set of
$K^0_S$ candidates further requirements are made:
\begin{itemize}
\item   Track pairs are required not to be 
        $\Lambda$ decay candidates. That is, it is required that
        the invariant mass of the track pair, $m_{p\pi}$, 
        satisfy $m_{p\pi} > 1.125\,$GeV/c$^2$, 
        where the higher momentum track is assumed 
        to be that of a proton and the lower that of a pion.
\item   The difference of the
        distance of closest approach to the primary 
        event vertex in the transverse plane, 
        $\Delta d$, 
        of the tracks making up
        the pair is required to satisfy
        $$|\Delta d| > \frac{1+10\exp[-p_T]}{5}\, \mbox{\rm cm}, $$
        where $p_T$ (in GeV/c) is the transverse
        momentum of the $V^0$ candidate.
        The distance of closest approach of a track, 
        $d$, is given a sign 
        that depends on the track's charge and 
        hence large values of $|\Delta d|$ result for  
        track pairs due to the decay of a $V^0$ into oppositely 
        charged particles at a vertex removed from the primary
        vertex. The applied restriction is a function of the
        $p_T$ to counteract the tendency of large $p_T$ to 
        cause small values of $|d|$.  
\item   It is required that $| \cos \theta^\star | < 0.95 $, where
        $\theta^\star$ is the angle in the transverse plane between
        the vector linking the primary and secondary vertices and the
        charged track direction in the rest frame of the track pair.
\end{itemize}

\noindent        
A set of potential $\Lambda$ decays is defined 
from the sample of $V^0$
candidates by demanding that:
\begin{itemize}
\item   Track pairs are not 
        $K^0_S$ decay candidates. That is, it is required that
        the invariant mass of the track pair, $m_{\pi\pi}$, 
        satisfy $m_{\pi\pi} < 0.45\,$GeV/c$^2$, 
        assuming both tracks are due to pions.
\item   Track pairs do not result from photon conversions. That is,
        the mass of the track pair $m_{ee}$, 
        assuming these to be electron and
        positron, is such that $m_{ee} > 0.05\,$GeV/c$^2$.
\item   The distance of closest approach of the tracks making up
        the pair is required to satisfy $|\Delta d| > 0.5\,$cm.
\end{itemize}

\begin{figure}[htbp]  \centering
\epsfig
 {file=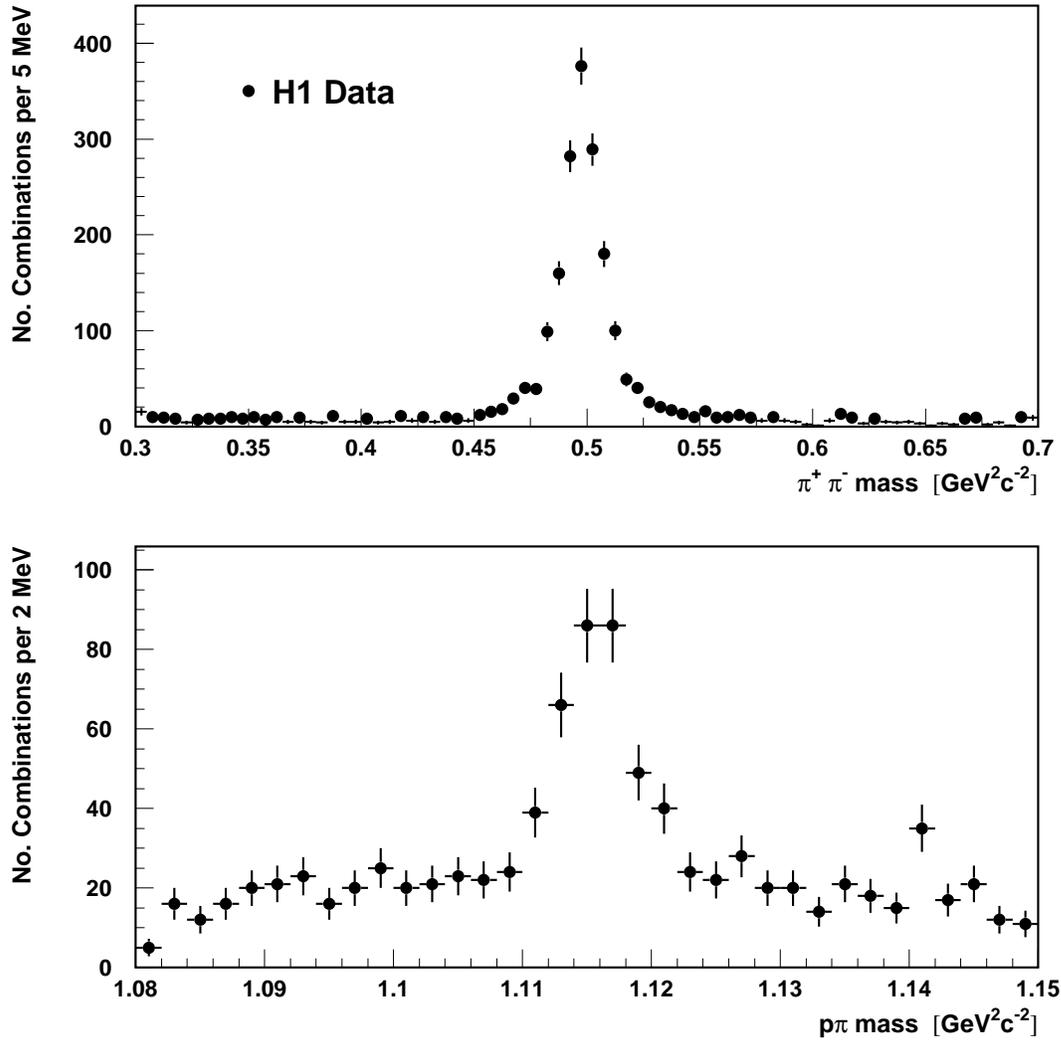,height=160mm,width=160mm,angle=0}
\caption{Pion-pion (upper graph) and
proton-pion (lower graph) mass spectra for $V^0$ candidates in the 
kinematic region 
$10 < Q^2 < 70\,$GeV$^2$, $10^{-4} < x < 10^{-2}$ 
and $0.05 < y < 0.6$
following the selection procedure described in the text.}
  \label{fig002}
\end{figure}

The invariant mass spectra of the track pairs for the
above samples
are shown in 
figure~\ref{fig002}.
These are calculated assuming that both tracks are due to pions
for the $K^0_S$ candidates and for the $\Lambda$ candidates 
that the higher momentum track is due to a proton,
that with lower momentum to a pion. The final step in the
selection of $K^0_S$ candidates is the
requirement that the invariant mass of the track pair,
$m_{\pi\pi}$, 
satisfy $0.45 < m_{\pi\pi} < 0.55\,$GeV/c$^2$. 
The $\Lambda$ candidates
are selected by requiring 
that $m_{p\pi}$ satisfy 
$1.1066 < m_{p\pi} < 1.1246\,$GeV/c$^2$.
The numbers of $K^0_S$ and $\Lambda$ candidates are
$1\,813$ and $395$ respectively. 

The masses of the $K^0_S$ and $\Lambda$ are extracted from
the spectra shown in figure~\ref{fig002}
by fitting a Gaussian to the obvious peaks, using various
functional forms to describe the remaining background.
The backgrounds to the $K^0_S$ and $\Lambda$ signals are 
$8 \pm 3\%$ and $29 \pm 7\%$ respectively.
The resulting masses are
$m_{K^0_S} = 497.9 \pm 0.2 \pm 0.2\,$MeV/c$^2$ and
$m_\Lambda = 1115.6 \pm 0.4 \pm 0.2\,$MeV/c$^2$, 
where the first error is 
statistical and the second systematic. 
The systematic error reflects the effects of
varying the assumptions on the functional form of the background
remaining under the signal and of residual uncertainties in the
calibration of the tracking system. 
These values are in agreement with the
current world averages quoted by the Particle Data Group 
(PDG)~\cite{PDG}, namely 
$m_{K^0}^{PDG}=497.672 \pm 0.031\,$MeV/c$^2$ and
$m_\Lambda^{PDG}=1115.683 \pm 0.006\,$MeV/c$^2$.

The numbers of $K^0_S$ mesons and 
$\Lambda$ baryons after background subtraction are
$1\,655 \pm 41$ and $237 \pm 30$ respectively. 
There are $116 \pm 21$
$\Lambda$ baryons and $126 \pm 20$ $\overline{\Lambda}$ anti-baryons. 
These latter numbers do not sum to give the total 
$\Lambda$ plus $\overline{\Lambda}$ yield above 
as they are obtained from independent fits with 
independent background subtractions.

The detection efficiencies for $K^0_S$ mesons and 
$\Lambda$ baryons in the given $p_T^2$ and $\eta$ ranges 
are determined using 
Monte Carlo techniques. 
The efficiency for 
$K^0_S$ detection is $25\%$ at a 
$p_T^2$ of $0.25\,($GeV/c$)^2$ and
rises to $40\%$ for a $p_T^2$ of $1\,($GeV/c$)^2$ and above.
The detection efficiency is independent of
pseudo-rapidity over the measured range. 
The $\Lambda$ detection efficiency is $12\%$ for all $\eta$ and $p_T^2$.
There is no significant difference 
between the $\Lambda$ and $\overline{\Lambda}$
detection efficiencies. Neither the $K^0_S$ nor the $\Lambda$ 
detection efficiencies depend significantly on $x$ or $Q^2$.
 
The $K^0_S$ and $\Lambda$ lifetimes obtained using the 
above event samples
are $\tau_{K^0_S} = 87.7 \pm 7.0 \pm 4.5\,$ps and  
$\tau_\Lambda = 330 \pm 130 \pm 30\,$ps respectively.
The lifetimes are determined from the proper
decay length distributions after subtracting
the backgrounds extracted using the mass spectra.
These are largest at small decay lengths.
A correction is also applied 
for the decrease in $K^0$ and $\Lambda$
detection efficiency
which occurs at small decay lengths.
This is determined using the
LEPTO Monte Carlo with default JETSET parameters. The 
systematic error reflects the uncertainties
in the background subtraction mentioned above and 
in the correction applied. The latter are estimated
using half the spread of results obtained when the correction
is derived using the LEPTO and CDM Monte Carlo programs with
default JETSET parameters, with an additional 
contribution to cover the effects of
deficiencies in the simulation of the 
H1 apparatus. The good agreement of the above results
with the PDG world averages, 
$\tau_{K^0_S}^{PDG} = 89.26 \pm 0.12\,$ps
and $\tau_\Lambda^{PDG} = 263.2 \pm 2.0\,$ps,
confirms that the background
subtraction and correction procedures are well understood.

\section{Results}
The observed numbers of $K^0_S$ mesons and $\Lambda$ baryons, 
after correction for event selection efficiency,
particle detection efficiencies and branching
ratios into the observed decay channels, 
correspond to 
the production of $n_{K^0} = 16\,414 \pm 403 \pm 709$ $K^0$ mesons
and $n_\Lambda = 2\,879 \pm 373 \pm 341$ $\Lambda$ baryons in the
pseudo-rapidity and transverse momentum range defined by
$-1.3 < \eta < 1.3 $ and 
$0.25 < p_T^2 < 4.5\,($GeV/c$)^2$ 
and in the kinematic region
$10 < Q^2 < 70\,$GeV$^2$, 
$10^{-4} < x < 10^{-2}$ and $0.05 < y < 0.6$. 
The corrections are calculated
by comparing the numbers of $K^0$ mesons 
and $\Lambda$ baryons in events  
generated using the LEPTO Monte Carlo with default 
JETSET parameters with the numbers 
found after a Monte Carlo simulation of the effects of the
H1 detector, reconstruction of the Monte Carlo data using the
same programs as used for the H1 data, and application of the 
selection criteria described above.
The systematic errors represent half the spread 
in the determination of the correction factors
arising from the use of the LEPTO and CDM generators with
default JETSET parameters,
the aforementioned uncertainties in the parameterisation
of the background mass spectra in the $K^0_S$ and $\Lambda$
mass regions, residual uncertainties in the calibration 
of the tracking system and deficiencies in the simulation of the
H1 apparatus in the Monte Carlo program used.
The effect of hard initial state photon radiation on these numbers, 
studied using DJANGO, is less than $2\%$. 
Hence no correction 
is applied, rather a contribution is added to the 
systematic error to cover the slight uncertainties 
introduced by
neglecting this effect. 

Monte Carlo simulations indicate that the above numbers correspond
to about $20\%$ of the total number of
$K^0$ mesons and the same proportion of all $\Lambda$
baryons produced in the $x$ and $Q^2$ region studied.  
The systematic 
uncertainties associated with the extrapolation from the measured
$\eta$ and $p_T^2$ region
to all $\eta$ and $p_T^2$ are large. 
Hence results quoted in this section are for the
restricted region in which accurate measurement is possible. 

\begin{figure}[htbp]  \centering
\epsfig
 {file=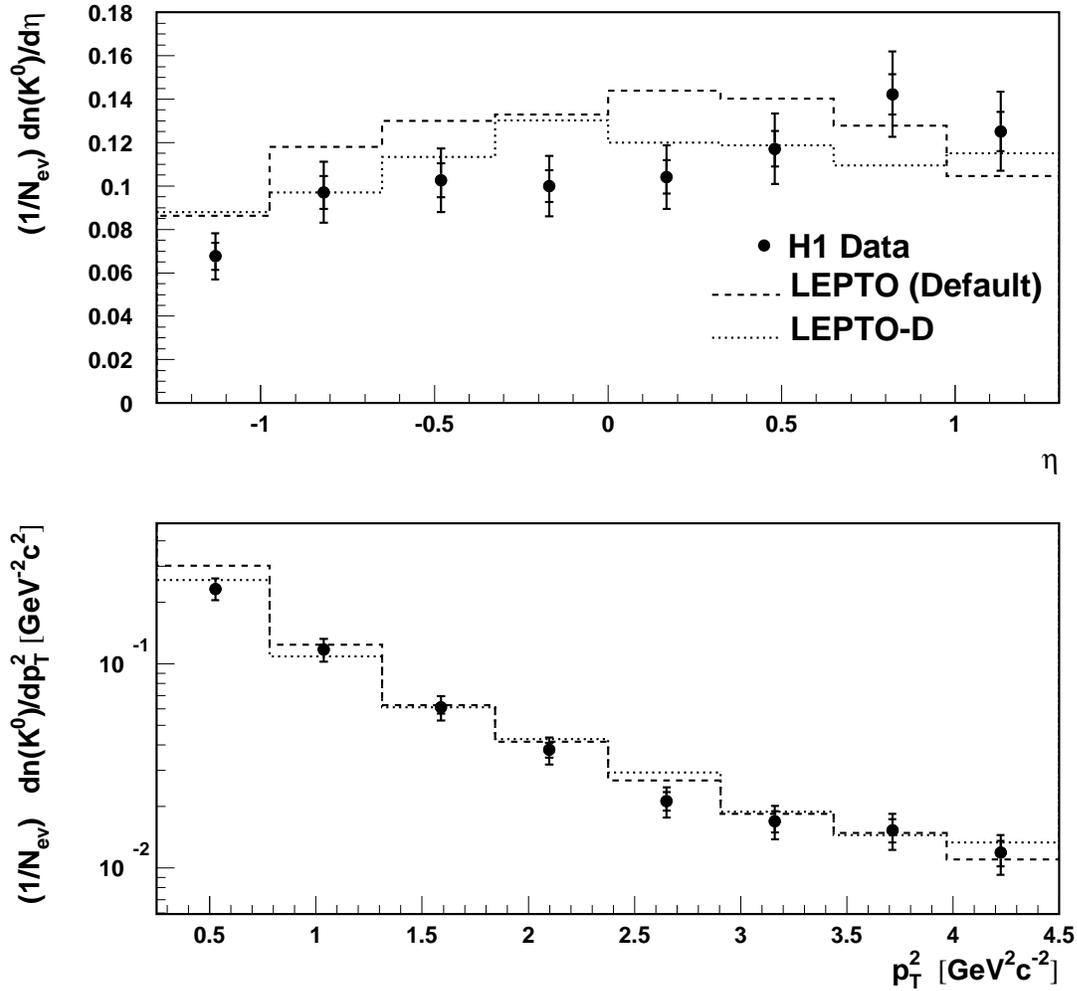,height=160mm,width=160mm,angle=0}
\caption{Measured 
$\eta$ spectrum (upper graph) for $K^0$ production 
in the range
$0.25 < p_T^2 < 4.5\,($GeV/c$)^2$  
and $p_T^2$ spectrum (lower graph) in the range 
$-1.3 < \eta < 1.3$ 
in non-diffractive DIS  
for the kinematic region
$10 < Q^2 < 70\,$GeV$^2$, $10^{-4} < x < 10^{-2}$ 
and $0.05 < y < 0.6$
compared with predictions from LEPTO with default JETSET parameters
(LEPTO (Default)) and
LEPTO with DELPHI JETSET parameters (LEPTO-D), see table 1.
}
  \label{fig003}
\end{figure}

For the reasons discussed in
the introduction, it is of interest to study 
the $K^0$ and $\Lambda$ spectra in $\eta$ and $p_T^2$.
The BFKL mechanism may 
cause excess strangeness production at large 
values of $\eta$ and $p_T^2$, whereas
instanton induced events contain
an excess of strange particles in the range 
$0 
\begin{array}{c}{\scriptstyle < } 
\\ [-0.35cm]{\scriptstyle \sim } \end{array} 
\eta
\begin{array}{c} {\scriptstyle < } 
\\ [-0.35cm] {\scriptstyle \sim } \end{array} 
3$. 
The relevant measured 
spectra for $K^0$ production in NDDIS,
in the kinematic region defined above,
are shown in 
figure~\ref{fig003}. 
This figure has been corrected for
acceptance and detection efficiency effects using 
Monte Carlo simulations, as have all subsequent
figures. Here and in all following figures in which 
two sets of error bars are displayed
the total errors are the result of 
adding in quadrature 
the statistical errors, shown using the inner error bars, and the 
systematic errors. 
The latter are estimated
as previously described.
Comparisons are made with the predictions of LEPTO using 
the default fragmentation parameters ($\lambda_s =0.3$), and 
using the set of parameters extracted by the DELPHI 
collaboration ($\lambda_s =0.23$, see table~\ref{tab001}). 
The normalisation
of the Monte Carlo predictions
is to the corrected number of 
NDDIS events in the specified $x$ and $Q^2$ range. 
In this and in other 
distributions it is observed that the description of the data 
provided using the DELPHI JETSET parameters
in the LEPTO and CDM Monte Carlo models 
is significantly better than that 
obtained with the default settings. The parameter change that has
the largest effect on the Monte Carlo predictions discussed here
is the modification to $\lambda_s$.  
A similar level of agreement is obtained if the
strangeness suppression factor 
favoured by E665 ($\lambda_s =0.2$) is used.
Once the modification to the fragmentation parameters
has been made the description of the data is
reasonable. There is no 
evidence for anomalous strangeness production
in either the absolute normalisation or the shape of the 
measured distributions.

The above agreement of the Monte Carlo predictions
for the average number of $K^0$ mesons per event
with the measurement does not prove that the
distribution of the number of $K^0$ 
mesons per event is correctly described.
The models might produce a larger number of 
strange particles in a smaller proportion of the 
events than observed in the data, or vice-versa. 
However, this is not
the case, as may be seen from the following observation.
The corrected proportion of NDDIS
events containing two kaons in the
kinematic region studied
is $0.044 \pm0.009 \pm 0.012$. 
This number is to be compared with the
LEPTO Monte Carlo predictions of $0.065$ and $0.056$ with 
default and DELPHI JETSET parameters respectively, and 
with the corresponding
CDM predictions of $0.060$ and $0.044$. 
HERWIG predicts that the proportion of di-kaon events is $0.059$.
Agreement with the 
Monte Carlo models which use JETSET is again 
observed to be better 
if the lower $\lambda_s$ value is used.
There is no evidence for anomalous production
of events containing many kaons. 

\begin{figure}[htbp]  \centering
\epsfig
 {file=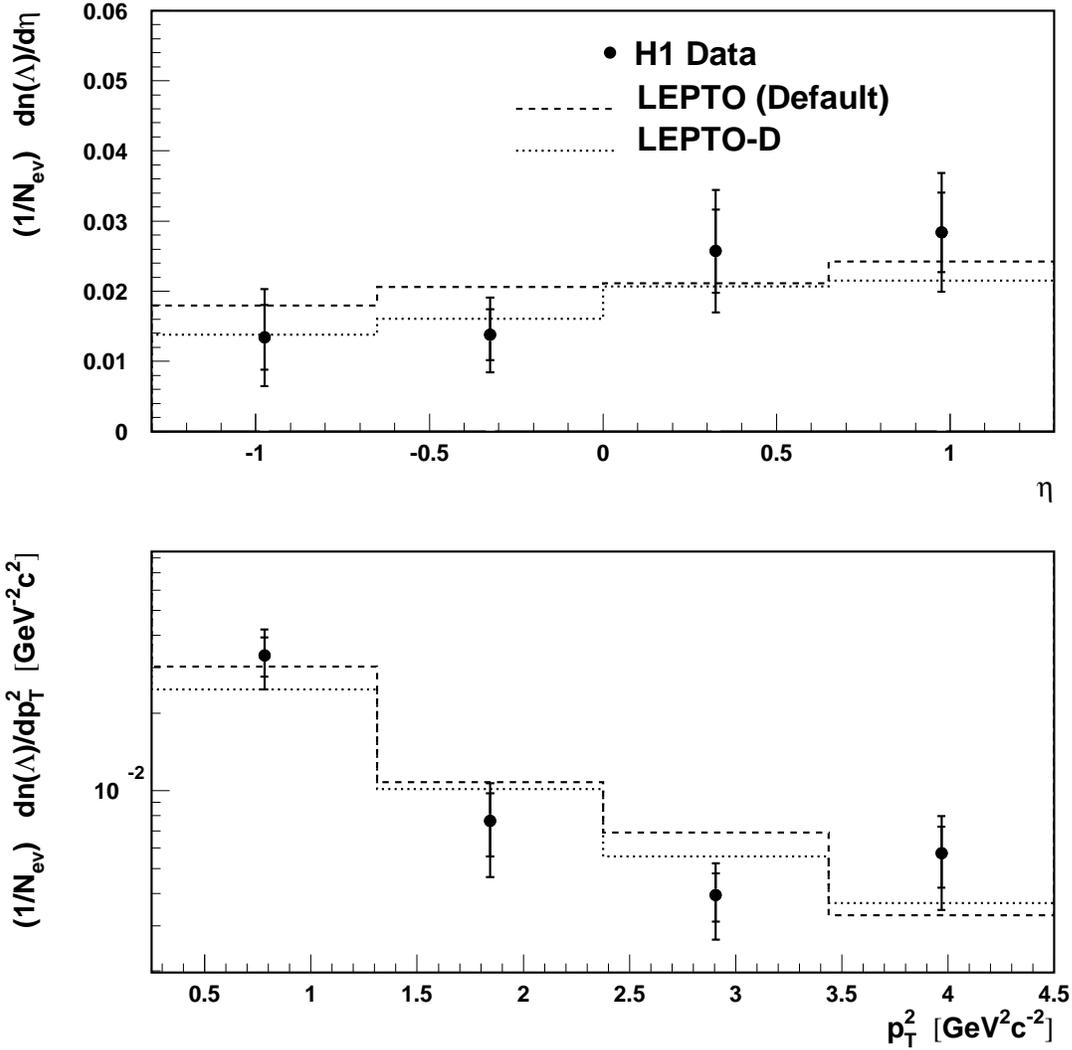,height=160mm,width=160mm,angle=0}
\caption{Measured $\eta$ spectrum (upper graph)
for $\Lambda$ production in the range
$0.25 < p_T^2 < 4.5\,($GeV/c$)^2$  
and $p_T^2$ spectrum (lower graph) in the range 
$-1.3 < \eta < 1.3$ 
in non-diffractive DIS  
for the kinematic region 
$10 < Q^2 < 70\,$GeV$^2$, $10^{-4} < x < 10^{-2}$ 
and $0.05 < y < 0.6$
compared with predictions from LEPTO with default JETSET parameters
(LEPTO (Default)) and
LEPTO with DELPHI JETSET parameters (LEPTO-D), see table 1.
}
  \label{fig004}
\end{figure}

The $\Lambda$ spectra in $\eta$ and $p_T^2$ are shown in
figure~\ref{fig004}. 
The statistical precision of these
data does not allow 
discrimination between the default and DELPHI sets of JETSET 
parameters. The broad agreement with the Monte Carlo
predictions indicates that the mechanisms 
responsible for di-quark
production in the fragmentation in 
DIS at low $x$ are similar to those at work in
other processes and at higher values of $x$. 

\begin{figure}[tbp]  \centering
\epsfig
 {file=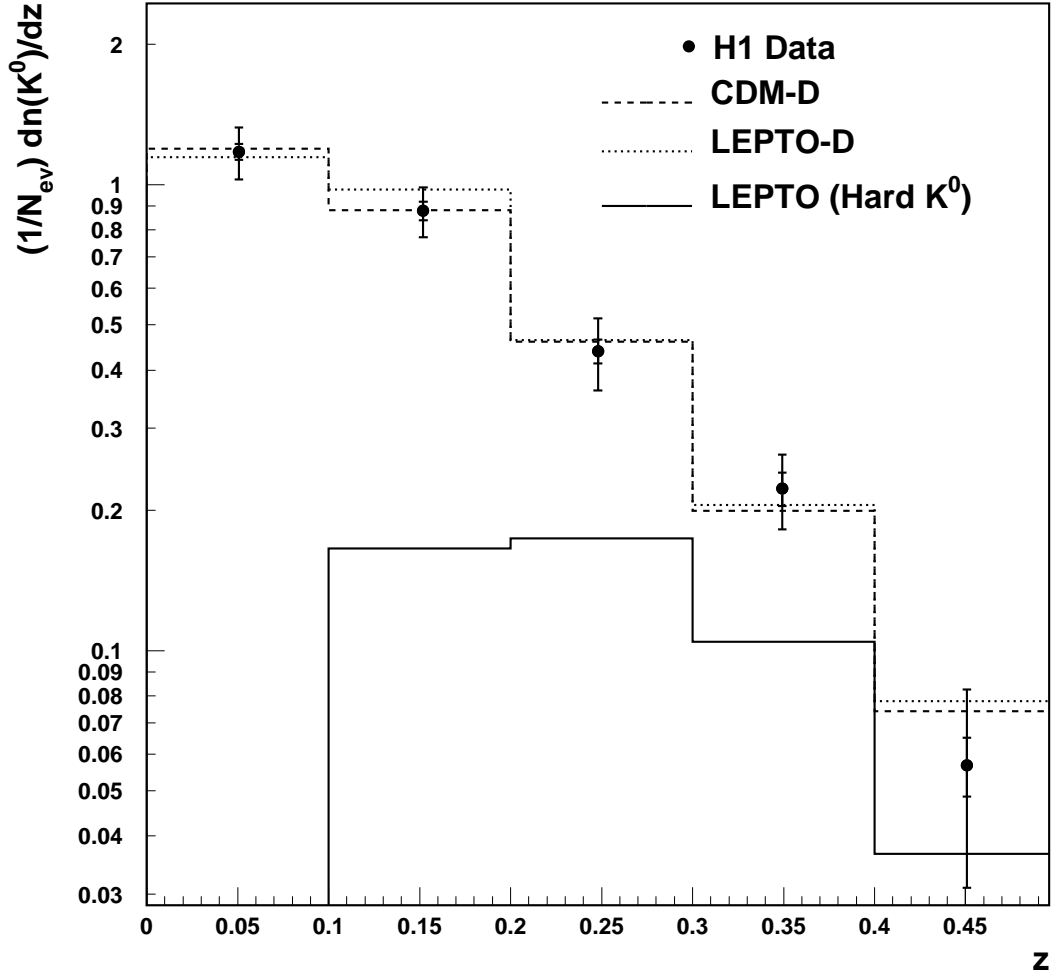,height=160mm,width=160mm,angle=0}
\caption{Measured 
$z$ spectrum for $K^0$ production in the range
$0.25 < p_T^2 < 4.5\,($GeV/c$)^2$ 
and $-1.3 < \eta < 1.3$
in non-diffractive DIS  
for the kinematic region  
$10 < Q^2 < 70\,$GeV$^2$, $10^{-4} < x < 10^{-2}$
and $0.05 < y < 0.6$
compared with predictions from LEPTO 
with DELPHI JETSET parameters (LEPTO-D) and the 
CDM with DELPHI JETSET parameters (CDM-D), see table 1;
also shown is the spectrum obtained from LEPTO 
if only $K^0$ mesons which contain a 
strange quark produced in the hard interaction are considered
(LEPTO (Hard $K^0$)). 
}
  \label{fig005}
\end{figure}

In order to allow some separation between strangeness produced in the
hard scattering   
and in the softer fragmentation process,  
$K^0$ production is also studied as a function of the 
fragmentation variable $z$, defined as
$$z=\frac{E_{K^0}+p_{L,K^0}}{2E_q},$$
where $E_{K^0}$ and $p_{L,K^0}$ are the energy of the $K^0$ and
the component of 
its momentum along the struck quark direction, given by
the four-momentum $q+xp$. This may be calculated using the 
measured parameters of the scattered positron and is the direction
expected for the scattered quark in the na\"\i ve Quark Parton Model. 
The quantity
$E_q$ is the energy expected for the struck quark, assumed massless,
using the same approximation.
Strange quarks may be produced in the hard scattering
either through the interaction of the virtual boson with a 
strange sea quark, or directly via the BGF process, or through the 
decay of $c$ quarks produced via BGF.
The $K^0$ mesons formed from strange quarks produced by these means,
referred to as ``hard $K^0$s'',
are likely to be at larger $z$ than
those produced in the fragmentation chain following, say, the
interaction of the virtual photon with a $u$ quark.
Hence anomalous
$K^0$ production at large $z$ is likely to be associated with
the hard sub-process, that at smaller $z$ with the hadronisation.
The $K^0$ spectrum in $z$ 
is shown in  
figure~\ref{fig005}. 
The Monte Carlo curves displayed in this figure are obtained
using LEPTO and the CDM. 
Of interest is the curve obtained when only hard $K^0$
mesons are considered. These are found to populate the
high $z$ region of the distribution, as expected, and 
form about $20\%$ of the total number
of $K^0$ mesons produced by the LEPTO generator
in the kinematic region studied when using the 
DELPHI JETSET parameters.  About $40\%$ 
of these result
from $c$ quark decays. 
There is no evidence
for anomalous production of strangeness associated with either the high
or low $z$ regions of the distribution. 

\begin{figure}[tbp]  \centering
\epsfig
 {file=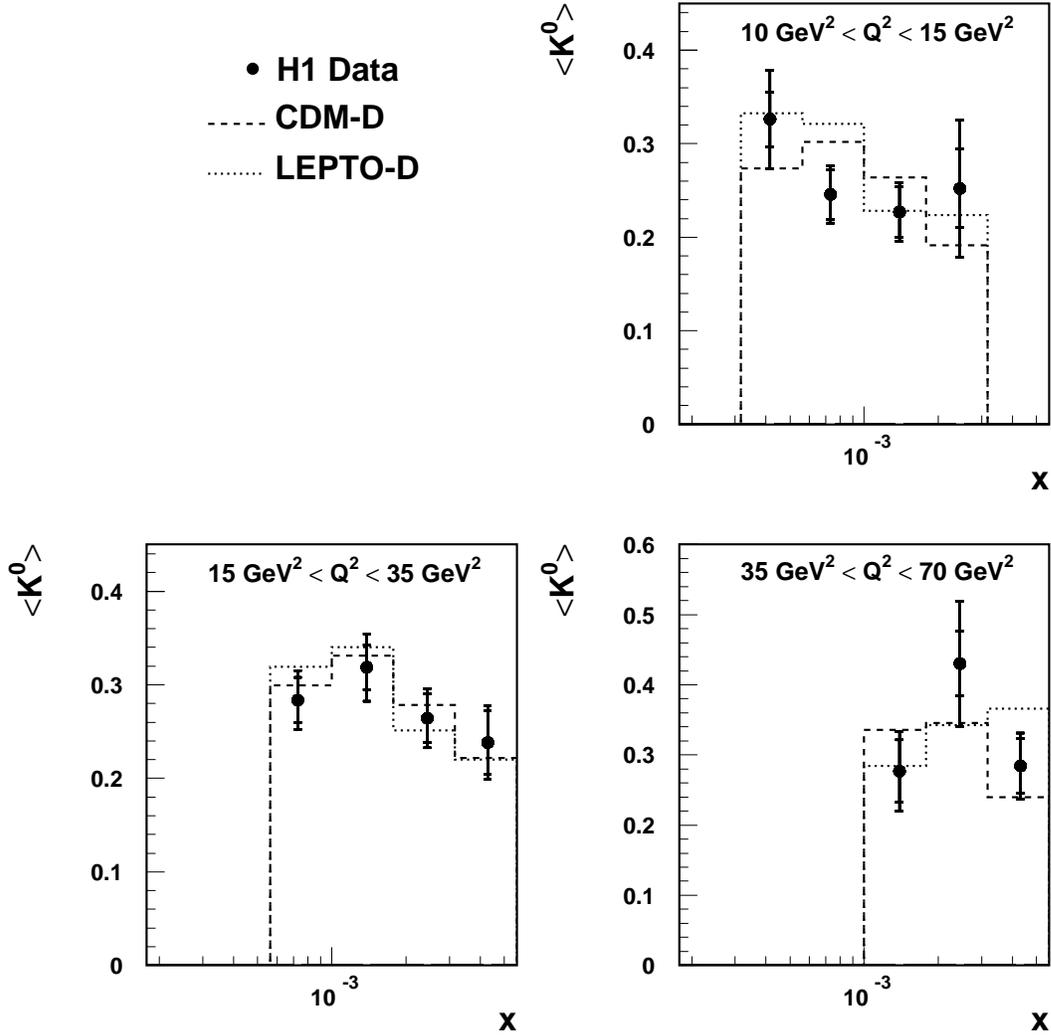,height=160mm,width=160mm,angle=0}
\caption{Mean numbers of 
$K^0$ mesons per NDDIS event produced in the range
$0.25 < p_T^2 < 4.5\,($GeV/c$)^2$ 
and $-1.3 < \eta < 1.3$
as a function of $x$, with the
restriction
$0.05 < y < 0.6$,
in non-diffractive DIS  
in the $Q^2$ regions indicated  
compared with predictions from
LEPTO with DELPHI JETSET parameters (LEPTO-D)
and the CDM with DELPHI JETSET parameters (CDM-D), see table~1.
}
  \label{fig006}
\end{figure}

In figure~\ref{fig006} 
are shown the mean numbers
of $K^0$ mesons observed per NDDIS event 
as a function 
of $x$ in the $Q^2$ regions
$10 < Q^2 < 15\,$GeV$^2$, 
$15 < Q^2 < 35\,$GeV$^2$ and
$35 < Q^2 < 70\,$GeV$^2$.
There is no evidence for any systematic deviation 
as a function of either $Q^2$ or $x$ between the
Monte Carlo predictions shown here and the data. 
As stated in the introduction, BFKL  
dynamics may  
cause additional $K^0$ production 
at large values of $\eta$ and at small $x$. 
It has been argued that the CDM gluon radiation prescription models
to some extent the effects of BFKL evolution 
on the hadronic state~\cite{CDMBFKL}. 
The similarity of the LEPTO and CDM predictions 
in figure~\ref{fig006} would
then indicate that, in the $\eta$ region studied, 
there is no sensitivity 
to any putative additional strangeness production due to 
BFKL dynamics. A definitive statement must await a Monte Carlo
implementation of the BFKL evolution prescription.

\begin{figure}[htbp]  \centering
\epsfig
 {file=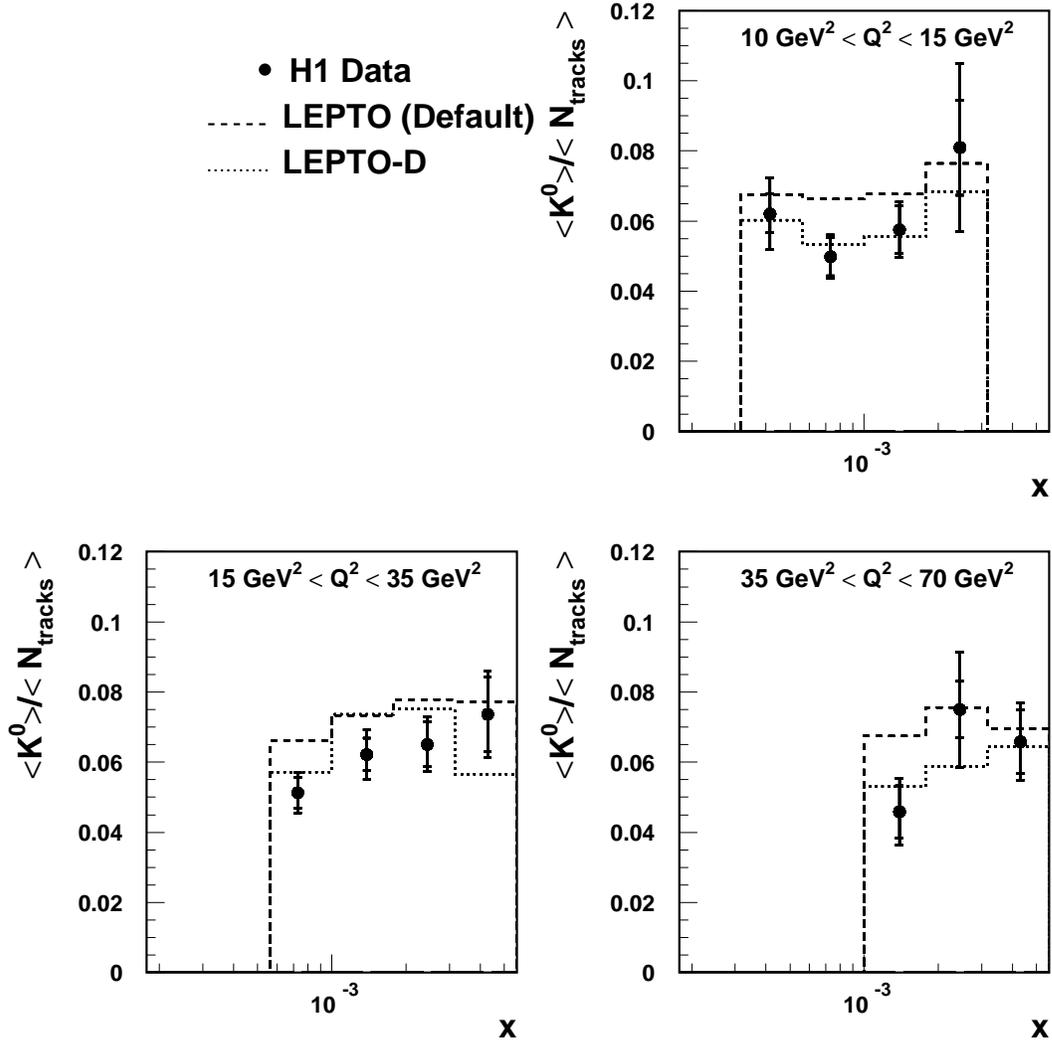,height=160mm,width=160mm,angle=0}
\caption{Measured ratios of the number of   
$K^0$ mesons in the range
$0.25 < p_T^2 < 4.5\,($GeV/c$)^2$ 
and $-1.3 < \eta < 1.3$
to all primary charged tracks
with $p_T > 0.15\,$GeV/c  
as a function of $x$, with the
restriction $0.05 < y < 0.6$
in non-diffractive DIS  
in the $Q^2$ regions indicated  
compared with predictions from LEPTO with default JETSET
parameters (LEPTO(Default)) and 
LEPTO with DELPHI JETSET parameters (LEPTO-D), see table~1.
}
  \label{fig007}
\end{figure}

An alternative way of examining the $K^0$ production data is to 
consider the ratio of the number of $K^0$ mesons produced to
the number of primary charged tracks. 
This latter is defined 
to be the number of 
primary charged tracks in the 
transverse momentum 
and pseudo-rapidity region given by
$p_T > 0.15\,$GeV/c, 
$-1.3 < \eta < 1.3 $ and is corrected for 
detection efficiencies using Monte Carlo simulations.
This means of presenting the data maintains sensitivity to effects 
that change the ratio of strange to non-strange quark production
while reducing the effect of global multiplicity changes. Hence,
for example,
shifts in the strangeness suppression factor may be observed 
with reduced sensitivity to hadronisation variables which tend
to change the overall track multiplicity.
The production ratios, in the $Q^2$ regions defined above, are 
shown in 
figure~\ref{fig007}. 
Again, better agreement with Monte Carlo simulations is observed 
when using the DELPHI than when using the default
JETSET parameters,
providing further evidence that it is indeed
the suppression of strangeness production in the 
hadronisation that is responsible for the differences, rather than
a global discrepancy in the modelling of the multiplicity.

\begin{figure}[htbp]  \centering
\epsfig
 {file=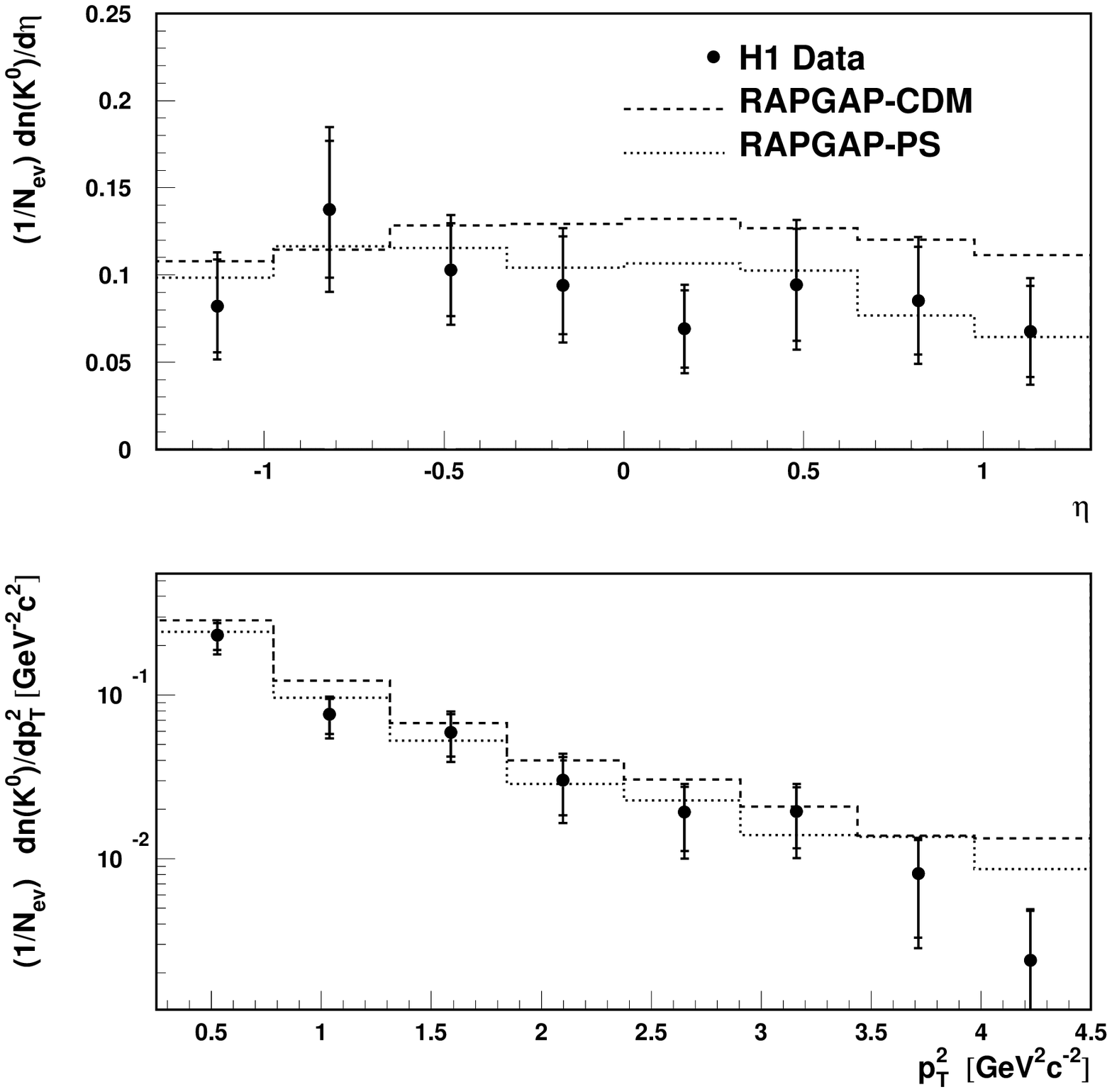,height=160mm,width=160mm,angle=0}
\caption{Measured 
$\eta$ spectrum (upper graph) for $K^0$ production in the range
$0.25 < p_T^2 < 4.5\,($GeV/c$)^2$ 
and $p_T^2$ spectrum (lower graph)
in the range               
$-1.3 < \eta < 1.3$ 
in diffractive DIS  
for the kinematic region 
$10 < Q^2 < 70\,$GeV$^2$, $10^{-4} < x < 10^{-2}$,
$0.05 < y < 0.6$ and $x_\pom < 0.05$
compared with predictions from RAPGAP-PS
and
RAPGAP-CDM.
}
  \label{fig008}
\end{figure}

Comparisons of the $K^0$ spectra in $\eta$ and $p_T^2$ for
DDIS events with calculations using RAPGAP are shown in
figure~\ref{fig008}. 
These data are corrected to the range $x_\pom < 0.05$ and to
the $x$ and $Q^2$ region defined above using the
results of the RAPGAP-PS Monte Carlo and a full simulation 
of the effects of H1 detector.  
The RAPGAP-CDM predictions, made with
the default JETSET parameters,
are seen to lie a little above the measurements, as is the 
case for NDDIS. The agreement with RAPGAP-PS is somewhat better.

The total rates of strange particle production in 
both DDIS and NDDIS are
listed in 
table~\ref{tab002}.
\begin{table}
\begin{center}
\begin{tabular}[htbp]{||l|l|l|l||}
\hline 
Data-Set & No. $K^0$/event &  No. 
$\Lambda$/event & No. $K^0$/No. tracks \\
\hline       
H1 NDDIS data & 0.278{$\pm0.008$}{$\pm0.012$} & 
                0.053{$\pm0.007$}{$\pm0.007$} & 
                0.058{$\pm0.002$}{$\pm0.003$}\\ 
\hline        
CDM (default)   & 0.316   & 0.052 & 0.066  \\
\hline
CDM (DELPHI)    & 0.284   & 0.047 & 0.062 \\
\hline
CDM (E665)      & 0.260   & 0.042 & 0.054 \\
\hline
LEPTO (default) & 0.320   & 0.055 & 0.068  \\
\hline
LEPTO (DELPHI)  & 0.282   & 0.047 & 0.064  \\
\hline
LEPTO (E665)    & 0.267   & 0.040 & 0.056  \\
\hline
HERWIG          & 0.316   & 0.047 & 0.068  \\
\hline
\hline
H1 DDIS data    & 0.238{$\pm0.026$}{$\pm0.016$}  & 
                  0.026{$\pm0.010$}{$\pm 0.025$} & 
                  0.063{$\pm0.007$}{$\pm0.004$}   \\ 
\hline
RAPGAP-PS       & 0.255   & 0.032 & 0.072 \\
\hline
RAPGAP-CDM      & 0.316   & 0.040 & 0.074 \\
\hline              
\end{tabular}
\caption{Measured mean $K^0$  
         and $\Lambda$ multiplicities with
         $0.25 < p_T^2 < 4.5\,($GeV/c$)^2$  
         and $-1.3 < \eta < 1.3$
         for the kinematic region  
         $10 < Q^2 < 70\,$GeV$^2$, $10^{-4} < x < 10^{-2}$ 
         and $0.05 < y < 0.6$
         and the ratio of the above multiplicities to the
         mean multiplicity of primary charged particles 
         with $p_T > 0.15\,$GeV/c in the same kinematic region
         in diffractive and non-diffractive DIS 
         together with the predictions
         of various Monte Carlo models.
}
  \label{tab002}
\end{center}
\end{table}
They reveal that the level of strangeness 
production in DDIS is somewhat
lower than that in NDDIS. This effect 
is probably due to the decreased 
overall level of particle production 
in DDIS in the $\eta$ region studied,
a consequence of the lower hadronic mass of the
fragmenting system ($M_X < W$),
rather than a difference in 
the strangeness suppression factor.
This argument is supported by
the observation that the ratio of $K^0$ mesons
to all charged particles in DDIS is consistent 
with that in NDDIS. 
Given that the overwhelming contribution to proton structure
in the $x$ range studied here is made by gluons, this observation 
is consistent with the electron-gluon scattering picture of
DDIS.
The consistency with the picture in which the scattering occurs between the
positron and a flavour singlet pomeron is demonstrated
by the agreement observed between the RAPGAP predictions
and the DDIS measurements of the $K^0$ to charged track ratio.

It is also apparent from 
table~\ref{tab002} that the difference in the 
two RAPGAP DDIS calculations of the
$K^0$ multiplicity largely disappears when
the ratio of that multiplicity to the 
primary charged particle multiplicity
is considered. This occurs as the
RAPGAP-CDM generator tends to produce a higher 
charged particle multiplicity 
than that seen with RAPGAP-PS.

\section{Comparison with Previous Measurements}
\begin{figure}[htbp]  \centering
\epsfig
 {file=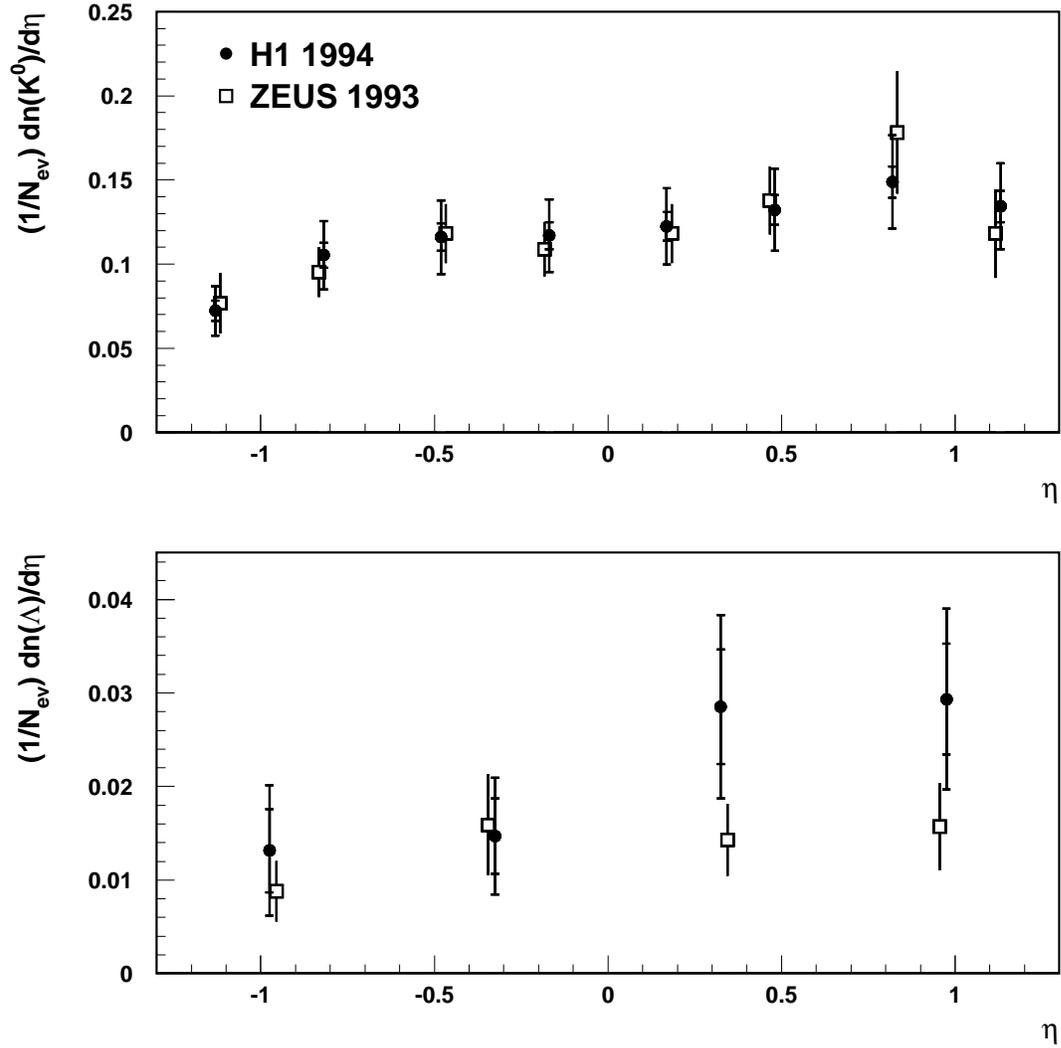,height=160mm,width=160mm,angle=0}
\caption{Measured 
$\eta$ spectrum for $K^0$ production (upper graph) 
and for $\Lambda$ production (lower graph) 
in the kinematic region
$10 < Q^2 < 640\,$GeV$^2$, $0.0003 < x < 0.01$ 
and $y > 0.04$; the $K^0$ results are for the range
$0.5 < p_T < 4.0\,$GeV/c and the $\Lambda$ results for  
$0.5 < p_T < 3.5\,$GeV/c, 
compared with the measurements of the ZEUS collaboration
made in deep-inelastic electron-proton scattering at HERA.
}
  \label{fig009}
\end{figure}

Figure~\ref{fig009} shows the $\eta$ spectra
for $K^0$ and $\Lambda$ production
in DIS (NDDIS and DDIS) 
after correction  
to the transverse momentum range 
$0.5 < p_T < 4.0\,$GeV/c in the case of the $K^0$ mesons and 
$0.5 < p_T < 3.5\,$GeV/c for the $\Lambda$ baryons  
and the kinematic
region $10 < Q^2 < 640\,$GeV$^2$, $0.0003 < x < 0.04$ 
and $y > 0.04$.
The correction is performed
using the LEPTO Monte Carlo with DELPHI JETSET parameters. The 
systematic errors shown include contributions due to the 
uncertainties in the correction procedure estimated using the spread
of results obtained when the default JETSET parameters are used
with LEPTO and both the default and DELPHI JETSET parameters
with the CDM. Although the central values 
obtained are very 
similar to those shown in figures~\ref{fig003} and~\ref{fig004},
the correction is to a kinematic
region which includes a significant
range outside the reach of the current measurement and hence
the associated systematic errors are larger than those 
shown in figures~\ref{fig003} and~\ref{fig004}.
These results are directly
comparable with the measurements of the 
ZEUS collaboration~\cite{ZK0}, 
also shown in figure~\ref{fig009}, obtained in deep-inelastic
electron-proton scattering at HERA. The two sets of data are
seen to be in reasonable agreement. 

In order to facilitate comparison with 
other measurements,
in the remainder of this
section the $K^0$ and $\Lambda$ NDDIS 
production results 
are presented as functions of various quantities in the hadronic
centre-of-mass frame with
the $z$ direction chosen to be that of the
exchanged virtual boson.
The transformation to this frame is made using the
measured energy and angle of the scattered electron.

\begin{figure}[htbp]  \centering
\epsfig
 {file=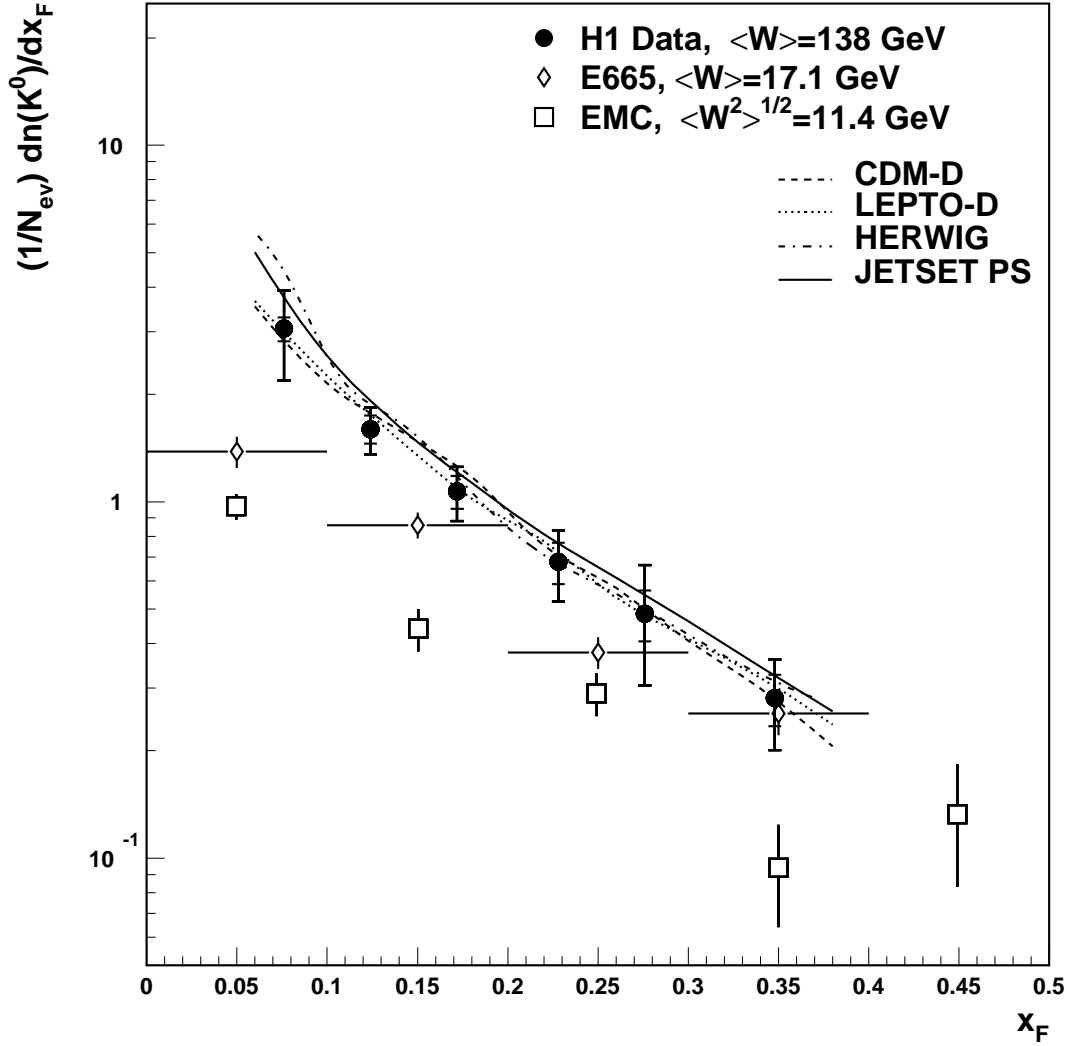,height=160mm,width=160mm,angle=0}
\caption{Production of $K^0$ mesons as a function 
of Feynman-$x$ ($x_F$) 
in the range
$10 < Q^2 < 70\,$GeV$^2$
compared with the 
muon-proton and muon-nucleon scattering results 
of the EMC and E665 collaborations,
the results of 
a Monte-Carlo simulation of electron-positron 
annihilation to $u$, $d$ and $s$ quarks
at the same centre-of-mass energy as the H1 data 
(JETSET PS) and various
other Monte Carlo simulations of DIS.
}
  \label{fig010}
\end{figure}

Figure~\ref{fig010} shows the $K^0$ production rate as a function
of Feynman-$x$,
the fractional longitudinal centre-of-mass momentum 
$x_F=2p^*_L/W$.
The data are also given in table~\ref{tab003}.
\begin{table}
\begin{center}
\begin{tabular}[htbp]{||c|c||}
\hline
 $x_F$  &  $\frac {1}{N} \frac {dn}{dx_F}$   \\
\hline
$0.05-0.10$ & $3.06\pm0.23\pm0.83$ \\
\hline
$0.10-0.15$ & $1.60\pm0.14\pm0.19$ \\
\hline
$0.15-0.20$ & $1.07\pm0.11\pm0.15$ \\
\hline
$0.20-0.25$ & $0.68\pm0.09\pm0.12$ \\   
\hline
$0.25-0.30$ & $0.48\pm0.08\pm0.16$ \\
\hline
$0.30-0.40$ & $0.28\pm0.05\pm0.07$ \\
\hline
\end{tabular}
\caption{The $K^0$ spectrum as a function of Feynman-$x$ ($x_F$) at
         $\langle W \rangle =138\,$GeV 
         for the range $10 < Q^2 < 70\,$GeV$^2$ (see figure~10).}
\label{tab003}
\end{center}
\end{table}
The $x_F$ range shown is restricted
to the region in which, for 
the laboratory pseudo-rapidity and transverse
momentum region over which particle identification is possible, 
the acceptance is at least $20\%$, namely
$0.05 < x_F < 0.40$. The average acceptance over this range is $35\%$.
The resulting mean centre-of-mass energy is  
$\langle W \rangle = 138\,$GeV. 
The measurements are corrected for the detector acceptance, the
region of laboratory pseudo-rapidity and transverse momentum in which 
particle identification is performed, and the kinematic region studied.
The correction is performed using 
the LEPTO Monte Carlo with DELPHI JETSET parameters. 
The systematic errors represent the dispersion
arising from using the default JETSET parameters
with LEPTO and both the default and DELPHI JETSET parameters
with the CDM in the 
correction procedure and the effects of the 
accuracy with which the transformation to the centre-of-mass
frame can be performed.
The shifts resulting from initial state photon radiation, studied
using DJANGO, are found to be small and hence
are not corrected for. The systematic error contains a contribution  
covering their effects.
Both the CDM and LEPTO Monte Carlo models describe the $x_F$ spectrum 
well, provided the DELPHI JETSET parameters are used. The agreement 
with the HERWIG Monte Carlo is also good.

Also shown in 
figure~\ref{fig010} 
are the data of the EMC collaboration~\cite{EMC},
obtained in muon-proton scattering at 
$\sqrt{\langle W^2 \rangle}=11.4\,$GeV,
and of the E665 collaboration~\cite{E665}, 
obtained in muon-nucleon scattering at  
$\langle W \rangle=17.1\,$GeV. 
These lie below the H1 data and fall less steeply with $x_F$,
the differences being more pronounced for the 
lower energy EMC data\footnote{Note that the EMC data here are at a
lower centre-of-mass energy than those compared with H1 results 
in~\cite{Eflow}.}.
This observation may be largely explained by the 
extra phase space available
for QCD radiation at the higher centre-of-mass energy 
available at HERA.

Comparisons of the $x_F$ spectra with electron-positron data are 
complicated by the direct production of heavy quarks
in electron-positron annihilation, with subsequent
decay to strange quarks and hence
enhanced strange particle production.
This problem is partially solved 
here by making the comparison
with Monte Carlo predictions for electron-positron annihilation
to $u$, $d$ and $s$ quarks only, at the H1 centre-of-mass energy.
Residual differences are 
to be expected as the proportion of primary $s$ quarks
produced in the simulation is still not the
same as that in DIS at this energy. 
The Monte Carlo used (JETSET with 
DELPHI tuning) accurately describes electron-positron 
annihilation data when the production of 
all possible flavours is allowed~\cite{DELPHI}.
A further complication is that 
the electron-positron annihilation data are available  
in terms of the fractional 
centre-of-mass momentum, $x_p=2p/W$. However, 
the transverse momenta in the hadronic centre-of-mass frame, $p_T^*$, 
of the $K^0$ mesons studied here
are generally much smaller than 
their longitudinal momenta and so the momentum  
$p\approx p_L$. Hence $x_F = x_p$
to an accuracy better than that represented by
the symbol size in the figures. 
The $K^0$ spectrum in $x_p$ from the JETSET simulation, 
divided by two to compensate for the production of 
a quark anti-quark pair in electron-positron annihilation, 
may thus be expected to be similar  
to the $x_F$ spectrum resulting from the
struck quark in NDDIS. 
That this is the case is seen in figure~\ref{fig010}.
 
\begin{figure}[htbp]  \centering
\epsfig
 {file=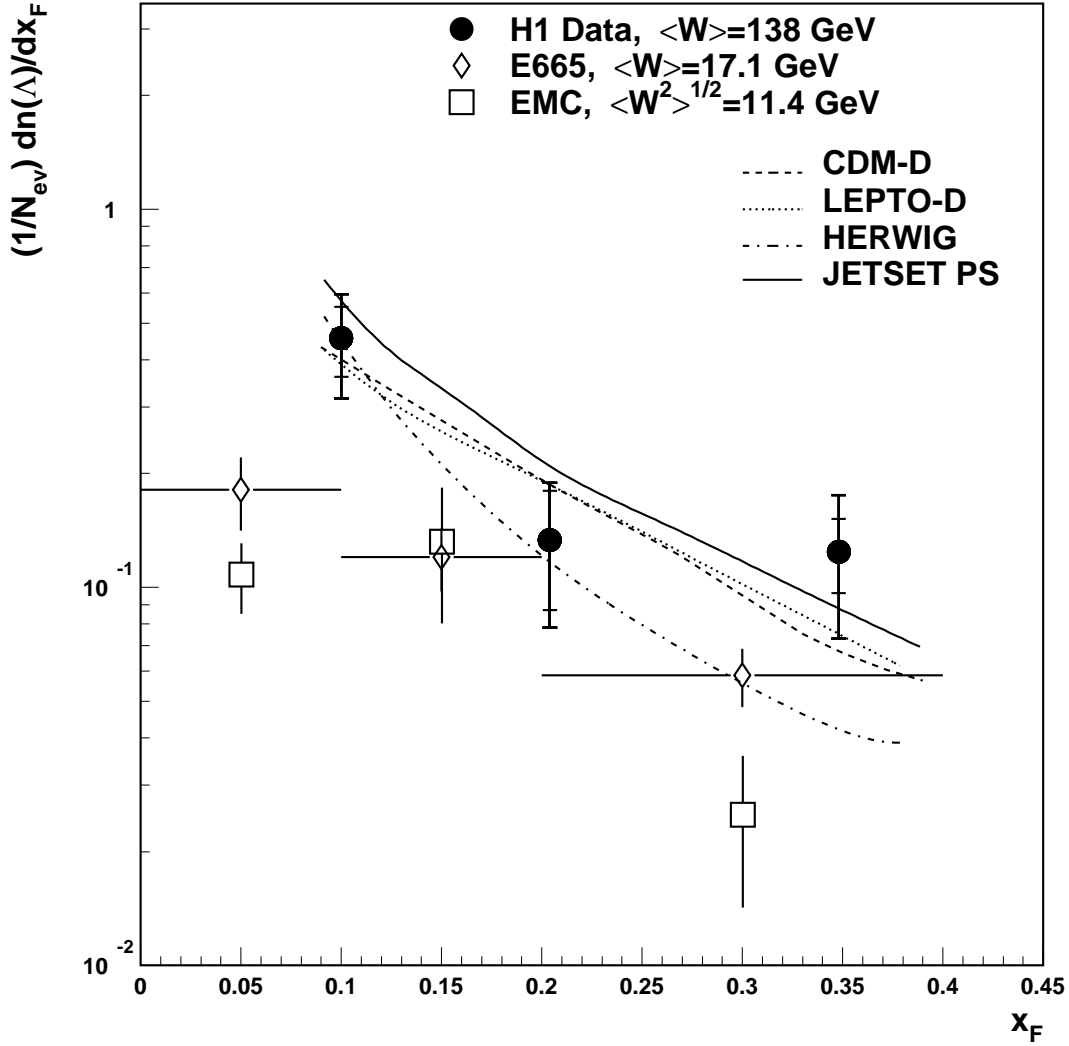,height=160mm,width=160mm,angle=0}
\caption{Production of $\Lambda$ baryons as a function of 
Feynman-$x$ ($x_F$) 
in the range $10 < Q^2 < 70\,$GeV$^2$
compared with the 
muon-proton and muon-nucleon scattering results 
of the EMC and E665 collaborations,
the results of a Monte-Carlo simulation of electron-positron 
annihilation to $u$, $d$ and $s$ quarks
at the same centre-of-mass energy as the H1 data 
(JETSET PS) and various
other Monte Carlo simulations of DIS.
}
  \label{fig011}
\end{figure}

Despite their lower precision, the $\Lambda$ measurements, shown in  
table~\ref{tab004} 
\begin{table}
\begin{center}
\begin{tabular}[htbp]{||c|c||}
\hline
 $x_F$  &  $\frac {1}{N} \frac {dn}{dx_F}$   \\
\hline
$0.05-0.15$ & $0.46\pm0.10\pm0.10$ \\
\hline
$0.15-0.25$ & $0.13\pm0.05\pm0.03$ \\
\hline
$0.25-0.45$ & $0.12\pm0.03\pm0.04$ \\
\hline
\end{tabular}
\caption{The $\Lambda$ spectrum as a function of Feynman-$x$ at
         $\langle W \rangle = 138\,$GeV 
         for the range $10 < Q^2 < 70\,$GeV$^2$(see figure~11).}  
  \label{tab004}
\end{center}
\end{table}
and figure~\ref{fig011}, 
reveal a similar trend to the $K^0$ results discussed above when 
compared with the lower energy EMC~\cite{EMC} and 
E665~\cite{E665} measurements. 
The agreement with the results of
a simulation of electron-positron annihilation
as described above is good. 
The HERWIG expectations differ from
those of LEPTO and the CDM somewhat, but the data are not precise
enough to discriminate between the models.

\begin{figure}[htbp]  \centering
\epsfig
 {file=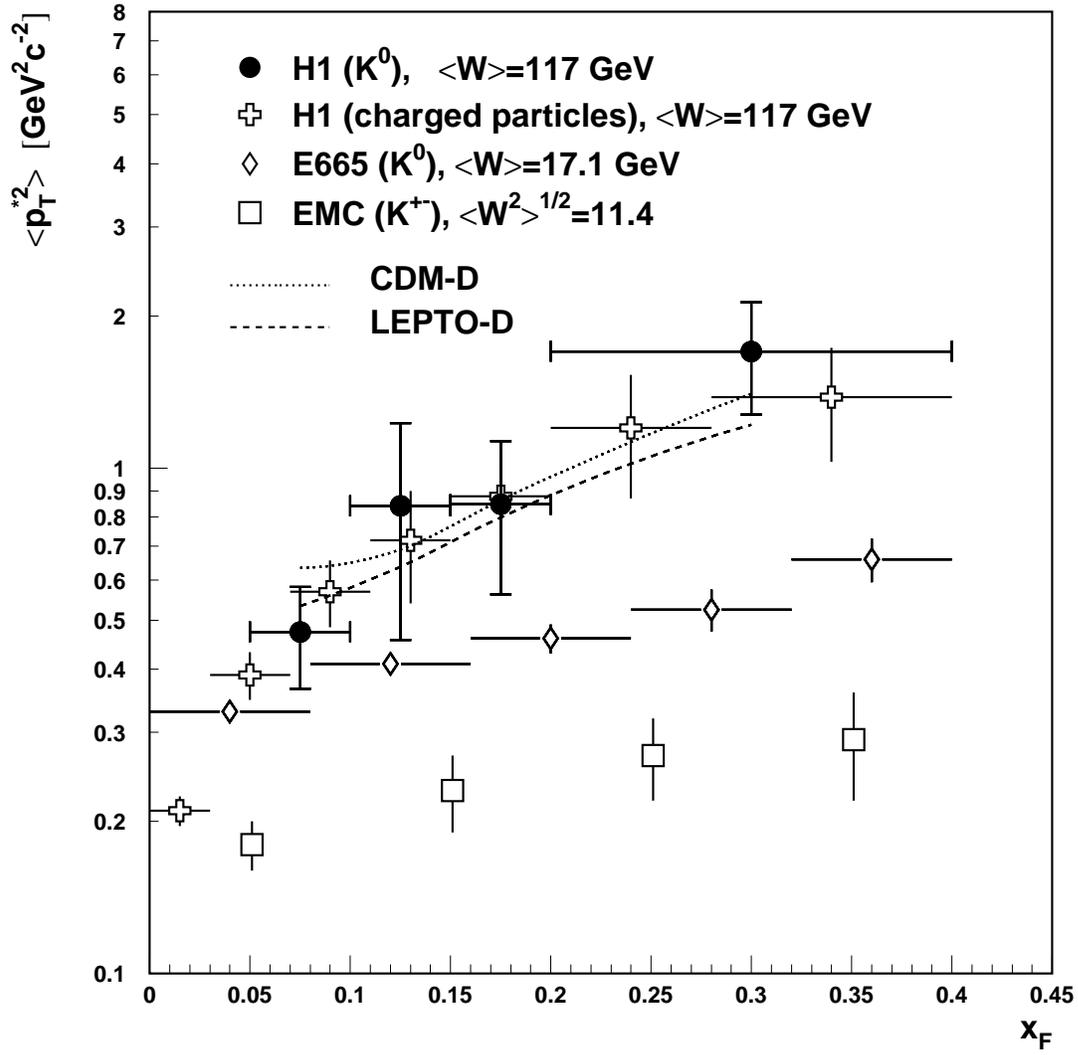,height=160mm,width=160mm,angle=0}
\caption{The mean squared transverse hadronic centre-of-mass
momentum ($\langle p_T^{*2} \rangle$) 
of $K^0$ mesons and of all charged particles
as a function of Feynman-$x$ ($x_F$)
in the range $10 < Q^2 < 70\,$GeV$^2$
compared with  
measurements of $K^0$ and $K^\pm$ production in muon-nucleon and
muon-proton scattering made by  
the E665 and EMC collaborations; also shown are 
various Monte Carlo predictions. 
}
  \label{fig012}
\end{figure}

The relationship between the fractional longitudinal momentum and the
mean squared transverse momentum, 
which is sensitive to QCD radiation effects,
is often illustrated in the ``seagull'' plot. Such a plot 
is shown for $K^0$ production in 
figure~\ref{fig012},  
the data also being given in 
table~\ref{tab005}.
%
%
\begin{table}
\begin{center}
\begin{tabular}[htbp]{||c|c||}
\hline
 $x_F$  &  $\langle p_T^{*2} \rangle$ \\
\hline
$0.05-0.10$ & $0.47\pm0.06\pm0.09$ \\   
\hline
$0.10-0.15$ & $0.84\pm0.06\pm0.38$ \\
\hline
$0.15-0.20$ & $0.85\pm0.09\pm0.27$ \\
\hline
$0.20-0.40$ & $1.70\pm0.08\pm0.42$ \\
\hline
\end{tabular}
\caption{Corrected values of $K^0$ $\langle p_T^{*2} \rangle$  
         as a function of Feynman-$x$ at  $\langle W \rangle =117\,$GeV
         for the range $10 < Q^2 < 70\,$GeV$^2$  
         (see figure~12).}
  \label{tab005}
\end{center}
\end{table}
%
%
%
%
Here, the $W$ range of the data is
restricted in order to reduce the mean
centre-of-mass energy  
to $\langle W \rangle = 117\,$GeV, comparable with 
previously published H1 data on charged particle 
production~\cite{Eflow}, also shown in the figure.
The $K^0$ and charged particle measurements 
are seen to be in good agreement and both lie 
at higher $\langle p_T^{*2} \rangle$ for a given $x_F$ than the lower
centre-of-mass energy measurements of the 
E665 and EMC collaborations.
This difference is expected within the framework of QCD
due to the effects of increased 
radiation at the higher energy. The CDM and LEPTO models, with
DELPHI JETSET parameters, are able to describe the data well.
HERWIG (not shown) describes the data 
shown here with a similar level of precision. 

\begin{figure}[htbp]  \centering
\epsfig
 {file=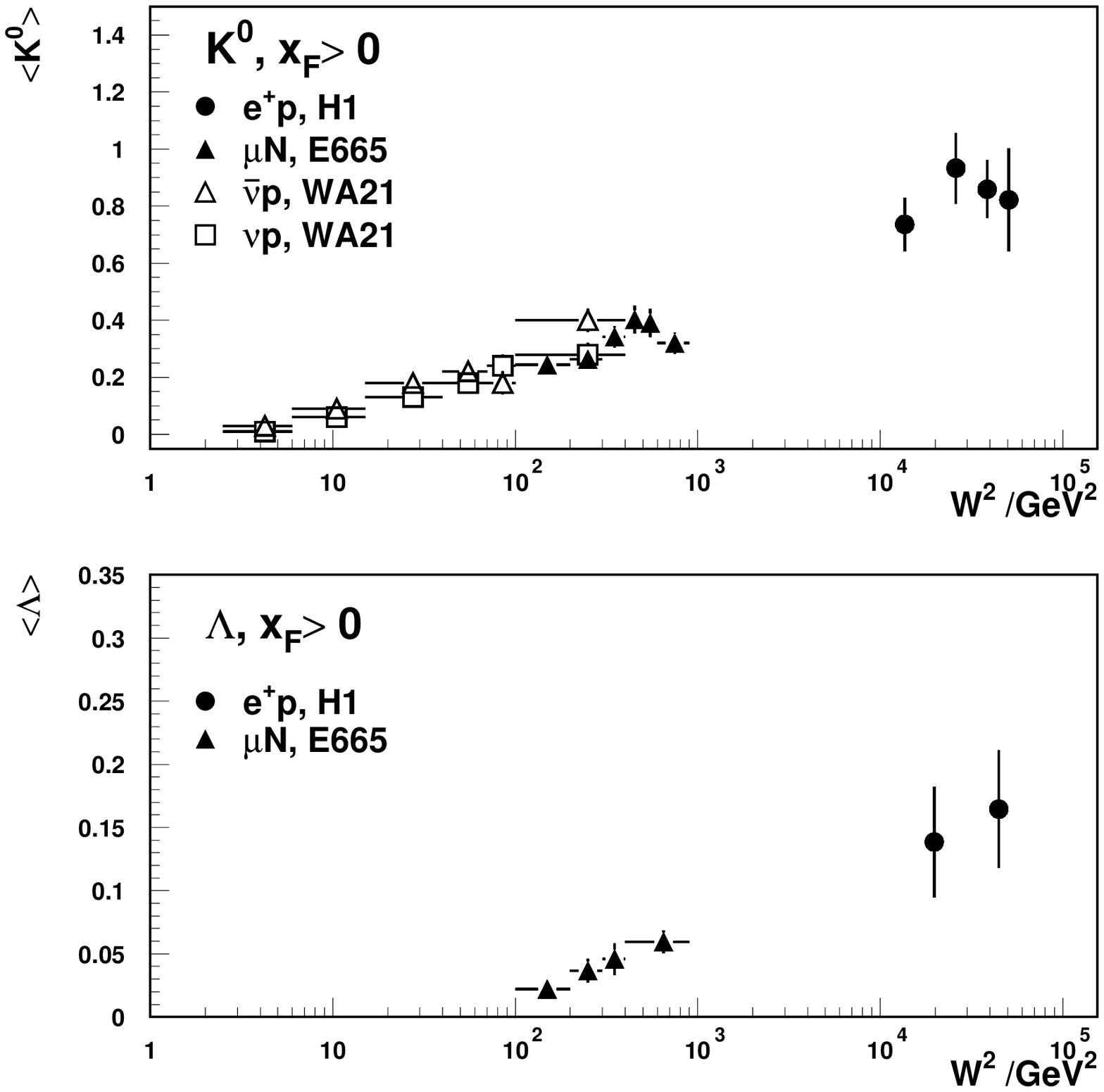,height=160mm,width=160mm,angle=0}
\caption{The mean $K^0$ (upper graph) and 
         $\Lambda$ (lower graph) multiplicities for $x_F > 0$
         as a function of $W^2$
         in the range $10 < Q^2 < 70\,$GeV$^2$ 
         compared with the 
         muon-nucleon scattering results 
         of the E665 collaboration and 
         WA21 measurements from 
         anti-neutrino and neutrino-proton scattering.
}
  \label{fig013}
\end{figure}

In order to study more directly the behaviour of $K^0$ and $\Lambda$
production with the hadronic centre-of-mass energy, the mean
$K^0$ and $\Lambda$ multiplicities are shown as a function of
$W^2$ in figure~\ref{fig013} and are also given in 
tables~\ref{tab006} and~\ref{tab007}.  
\begin{table}
\begin{center}
\begin{tabular}[p]{||c|c||}
\hline
 $W^2\,$(GeV$^2$) &  $\langle K^0 \rangle$ \\
\hline
$6400-19300$  & $0.74\pm0.03\pm0.09$ \\
\hline
$19300-32200$ & $0.93\pm0.05\pm0.11$ \\
\hline
$32200-45100$ & $0.86\pm0.06\pm0.08$ \\
\hline
$45100-58000$ & $0.82\pm0.09\pm0.16$ \\
\hline
\end{tabular}
\caption{Mean $K^0$ multiplicity 
         for $x_F>0$ in the hadronic centre-of-mass frame 
         as a function of $W^2$ (see figure~13).}
  \label{tab006}
\end{center}
\end{table}
\begin{table}
\begin{center}
\begin{tabular}[p]{||c|c||}
\hline
 $W^2\,$(GeV$^2$) &  $\langle \Lambda \rangle $ \\
\hline
$6400-15625$  & $0.13\pm0.03\pm0.03$ \\
\hline
$15625-58000$ & $0.17\pm0.03\pm0.04$ \\
\hline
\end{tabular}
\caption{Mean $\Lambda$ multiplicity for
         $x_F>0$ in the hadronic centre-of-mass
         system as function of $W^2$ 
         for the range $10 < Q^2 < 70\,$GeV$^2$(see figure~13).}
  \label{tab007}
\end{center}                          
\end{table}
These multiplicities are
corrected so that they represent the numbers of particles 
produced with $x_F > 0$.
A procedure similar to that described above
is used to obtain the correction and the associated systematic
error. Both the $K^0$ and $\Lambda$
measurements show that the logarithmic 
increase in mean multiplicity observed at lower energies (here
data are shown from the E665 collaboration and from the anti-neutrino
and neutrino-proton 
scattering experiments of the WA21 collaboration~\cite{WA21b})
persists to the much higher energies available at HERA.

\section{QCD Instanton Production}
\begin{figure}[htbp]  \centering
\epsfig
 {file=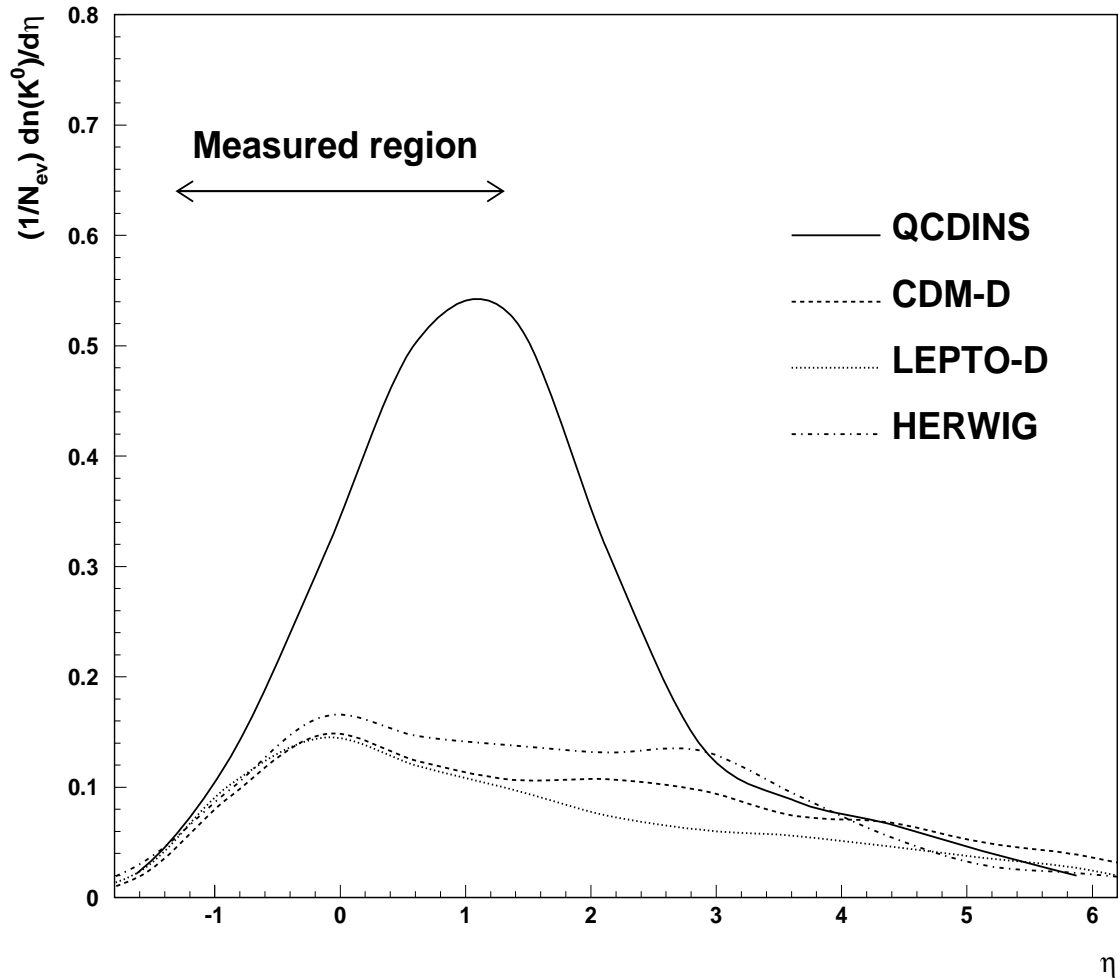,height=140mm,width=160mm,angle=0}
\caption{Numbers of $K^0$ mesons per event with
$0.25 < p_T^2 < 4.5\,($GeV/c$)^2$ 
as a function of $\eta$
for the kinematic region 
$10 < Q^2 < 70\,$GeV$^2$, $10^{-3} < x < 10^{-2}$
and $0.1 < y < 0.6$
predicted by models of non-diffractive DIS 
(broken lines) and by QCDINS (continuous line).
}
  \label{fig014}
\end{figure}

\begin{figure}[htbp]  \centering
\epsfig
 {file=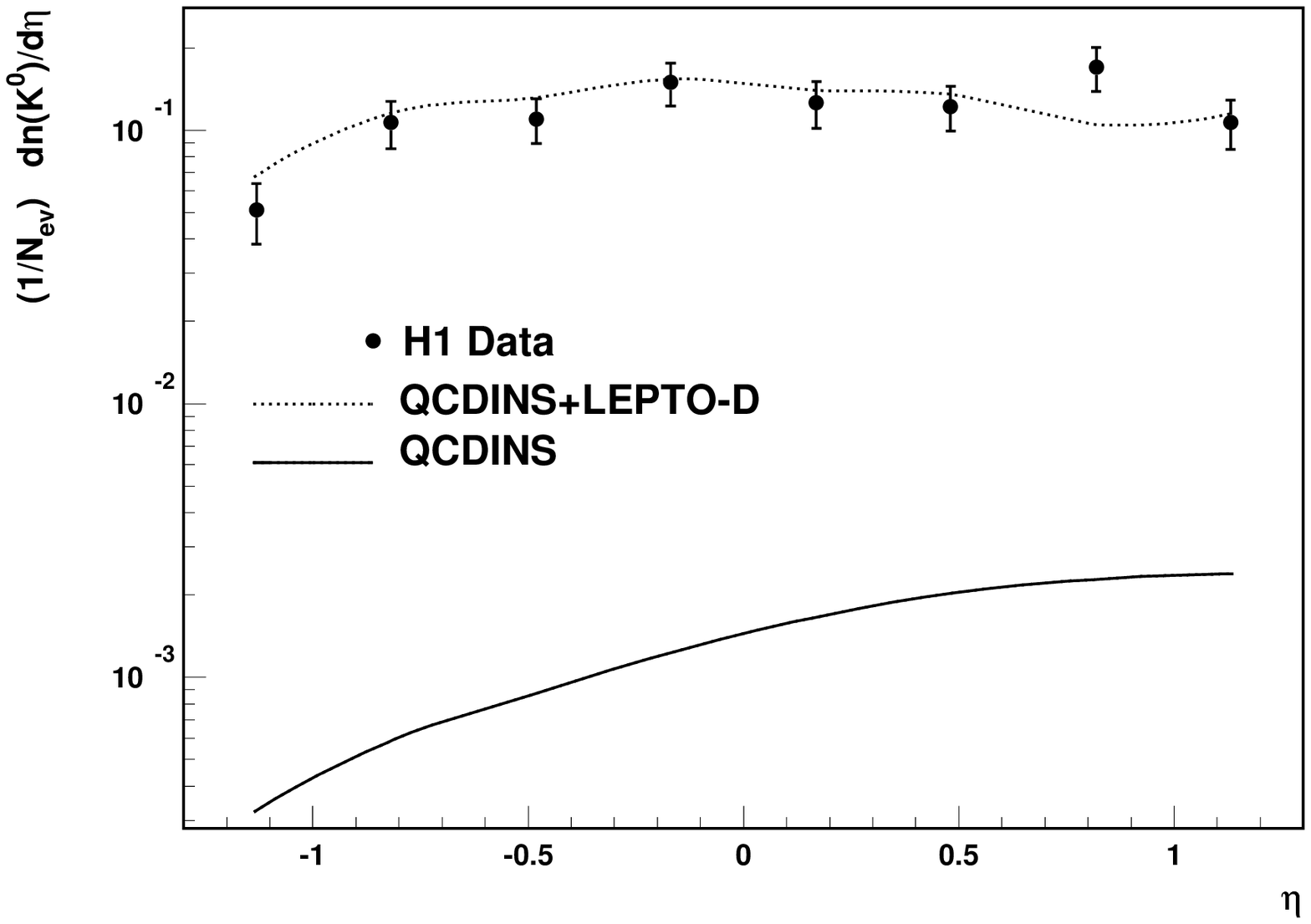,height=120mm,width=160mm,angle=0}
\caption{Measured numbers of $K^0$ mesons per event with
$0.25 < p_T^2 < 4.5\,($GeV/c$)^2$ 
for the kinematic region 
$10 < Q^2 < 70\,$GeV$^2$, $10^{-3} < x < 10^{-2}$
and $0.1 < y < 0.6$
(points) with the fit result (dotted line) and the
fraction of instanton induced 
events (solid line), $f=0.006$,
obtained as described in the text;
the errors shown are the statistical 
and systematic errors added in quadrature.
}
  \label{fig015}
\end{figure}

If a significant proportion of DIS events 
were induced by QCD instantons, large changes in the
strangeness composition of the hadronic final state would be 
expected. This is illustrated in 
figure~\ref{fig014} where, 
to ensure reliable simulation 
using the instanton generator, the kinematic 
region studied is further restricted to
$10 < Q^2 < 70\,$GeV$^2$, $10^{-3} < x < 10^{-2}$
and $0.1 < y < 0.6$. The number of events 
in this region is $16\,486$ 
after the application of the NDDIS selection criteria.
In order to determine an upper limit for the 
cross-section for instanton induced events
it is then assumed that the 
measured rate of $K^0$ production,
shown as a function of $\eta$ in figure~\ref{fig015}, is the
result of a proportion $f$ of instanton induced events with the rest
being due to the NDDIS models discussed above. A 
$\chi^2$ minimisation procedure 
is used to determine the value of $f$.

When using LEPTO and the CDM
to calculate the
rate expected for NDDIS without an instanton contribution, 
the DELPHI $\lambda_s $ value and other fragmentation parameters are
used as these are determined 
in electron-positron annihilation, in which
there is no contribution from instanton induced events.
The results obtained are $f=-0.044\pm0.032$ and 
$f=-0.002\pm0.031$ using QCDINS with HERWIG and the CDM respectively.
The largest value of $f$ is obtained using QCDINS with
LEPTO and is $f=0.006 \pm 0.030$.
The fit to the measured $K^0$ production rate 
resulting in this latter case is shown in 
figure~\ref{fig015}.
The $f$ values are consistent with being zero
and hence a
$95\%$ confidence level upper limit 
of $0.9\,$nb is placed
on the cross-section for instanton induced events
in the above kinematic region, where the
largest $f$ value is used
in the determination of the limit. The limit is of the same order 
of magnitude as a recent, albeit uncertain, determination of the
cross-section for QCD instanton induced events in DIS~\cite{Rome}. 

\section{Conclusions}
Strangeness production in deep-inelastic 
positron-proton scattering (DIS) is studied using the H1
apparatus at the HERA collider at DESY. 
The rates of
$K^0$ meson and $\Lambda$ baryon production are measured
in the kinematic region
$10 < Q^2 < 70\,$GeV$^2$, $10^{-4} < x < 10^{-2}$ 
and $0.05 < y < 0.6$.
No evidence for anomalous sources
of strangeness is observed. A reduction of the 
strangeness suppression factor, 
$\lambda_s $, from the current default
value in the JETSET string hadronisation scheme, $0.3$, to a value
of $0.23$ or $0.2$ significantly
improves the description of the 
$K^0$ measurements when using
both the LEPTO and CDM Monte Carlo models. This observation is
in agreement with the results of previous measurements  
made in DIS~\cite{E665,ZK0} 
and electron-positron annihilation~\cite{DELPHI}. 
The agreement 
with the latter suggests that the colour forces involved
in the hadronisation process are similar in electron-positron 
annihilation and DIS in the low $x$ region studied here.
Further evidence for this is the reasonable description of
the measured $\Lambda$ production rate 
given by Monte Carlo models using JETSET hadronisation.
Such agreement has also been observed in
electron-positron annihilation~\cite{DELLam}, 
suggesting that the
di-quark production mechanisms are similar in both processes.

No evidence is found
for differences in the relative 
rate of strangeness production
in diffractive as opposed to non-diffractive DIS, an observation 
consistent with both the hypothesis that these events result 
from deep-inelastic positron-pomeron scattering and the hypothesis
that they are the product of boson-gluon fusion reactions 
with subsequent modification of the quark and anti-quark colour
charges such that these form a colour singlet.

Direct comparisons of the fractional longitudinal hadronic 
centre-of-mass momentum ($x_F=2p_L^*/W$) spectra with
previous studies of $K^0$ and $\Lambda$ production 
in DIS~\cite{EMC,E665} at lower energy reveal an increase in the 
number of particles, particularly at 
low $x_F$, as a function of energy. This is 
expected within the framework of QCD 
and may be explained
as being due to the increased amount of radiation
at the higher energy.
The $x_F$ spectra are also compared with $x_p=2p/W$ spectra
obtained from a simulation of electron-positron annihilation 
to $u$, $d$ and $s$ quarks at the same centre-of-mass energy. These
are in good agreement, providing further evidence that the
processes involved in the hadronisation in DIS
at the low values of $x$ accessible at HERA
are the same as those in electron-positron annihilation.
A further consequence of the increase in QCD radiation
with energy is observed when the mean squared 
transverse momentum of the $K^0$ mesons in the 
hadronic centre-of-mass frame
($\langle p_T^{*2} \rangle $)
is studied as a function of $x_F$. 
The high energy data are seen to be at higher 
$\langle p_T^{*2} \rangle $ for a given $x_F$ than the lower energy 
data~\cite{E665,WA21b}. These observations 
are described well by models 
of DIS which incorporate QCD effects. 

As no evidence for anomalous levels of strangeness production is
found, an upper limit of $0.9\,$nb at the $95\%$ confidence level is placed
on the cross-section for QCD instanton induced events in DIS. 

\section{Acknowledgements}
We are grateful to the HERA machine group whose outstanding efforts
have made this experiment possible. We appreciate the immense effort
of the engineers and technicians who constructed and maintain the H1
detector. We acknowledge the support of the DESY technical staff and 
thank the DESY directorate for the hospitality extended to the 
non-DESY members of the collaboration.

\end{document}